\documentclass[aps,prl,twocolumn,superscriptaddress,10pt]{revtex4-2}
\usepackage{graphicx}
\usepackage{amsmath} 
\usepackage{bm}      % 新增这一行，专门用于粗体数学符号
\usepackage{amssymb}
\usepackage[utf8]{inputenc}
\usepackage{color}
\usepackage{amsmath}
\usepackage{amssymb}
\usepackage{xcolor} 
\usepackage{graphicx}
\usepackage{esint}
\usepackage[colorlinks=true,citecolor=blue!60!black,linkcolor=blue!60!black,urlcolor=blue!60!black]{hyperref}
\usepackage{appendix}
\usepackage{comment}
\makeatletter
\@ifundefined{textcolor}{}
{%
	\definecolor{red}{rgb}{0.75, 0.1, 0.1}
	\definecolor{BLACK}{gray}{0}
	\definecolor{WHITE}{gray}{1}
	\definecolor{GREEN}{rgb}{0,1,0}
	\definecolor{BLUE}{rgb}{0,0,1}
	\definecolor{CYAN}{cmyk}{1,0,0,0}
	\definecolor{MAGENTA}{cmyk}{0,1,0,0}
	\definecolor{YELLOW}{cmyk}{0,0,1,0}
}

\makeatother
\usepackage{mathtools}
\newcommand{\Tr}{\operatorname{Tr}}

\newcommand{\Lind}{\mathcal{L}}
\newcommand{\vp}{\mathbf p}

\begin{document}
\title{Reset-Enhanced Dissipative Quantum Metrology}

\author{Silan Li}
\affiliation{School of Physics and Electronics, Hunan University, Changsha 410082, China}

\author{Qingjun Tong}
\email{Corresponding author: tongqj@hnu.edu.cn}
\affiliation{School of Physics and Electronics, Hunan University, Changsha
	410082, China}

\author{Ruicheng Bao}
\email{Corresponding author: ruicheng@g.ecc.u-tokyo.ac.jp}
\affiliation{Department of Physics, Graduate School of Science,The University of Tokyo, Hongo, Bunkyo-ku, Tokyo 113-0033, Japan}

\begin{abstract}
Quantum metrology is ultimately shaped not only by measurements, but also by the dissipative dynamics that prepare and maintain the probe.
Nevertheless, a general  mechanism for enhancing both steady-state and transient precision in open quantum systems has been lacking.
Here we show that adding a simple reset channel provides such a mechanism: reset acts as a minimal dissipative control that can enhance metrological precision without modifying the intrinsic Hamiltonian or applying continuous coherent driving.
We derive a unified quantum-Fisher-information (QFI) reset-rate response identity, showing how reset enhances steady-state and transient estimation while revealing a sharp transient distinction.
For static-parameter estimation in diagonal dynamics, reset leaves the leading short-time QFI unchanged; for the separate task of elapsed-time estimation, it generates an immediate drift contribution that can scale quadratically with the reset rate in the short-time limit. For frequency estimation, the enhancement is instead dominated by a genuinely quantum eigenvector contribution.
These results establish reset as a simple dissipative resource for open-system quantum metrology.

\end{abstract}

\maketitle

\paragraph*{Introduction.---}
Quantum metrology converts state distinguishability into estimation precision~\cite{braunstein1994statistical,giovannetti2006quantum,paris2009quantum,toth2014quantum,liu2024fully,mukhopadhyay2025current}. 
It underlies sensing protocols across atomic, condensed-matter, photonic, and nanoscale platforms~\cite{giovannetti2011advances,neumann2013high,polino2020photonic,mitchell2004super,ding2022enhanced,pezze2007phase,gross2010nonlinear,ono2013entanglement,leibfried2004toward,bonato2016optimized,wang2019heisenberg,mihailescu2024quantum,petropoulos2024nanoscale,zhou2025entanglement,kuffer2025sensing,rosenberg2026witnessing,cao2026cross,hui2026ultrathin}.
Previous studies have focused on enhancing metrological precision under closed-system unitary dynamics~\cite{boixo2007generalized,roy2008exponentially,giovannetti2006quantum}.
In realistic sensors, however, the probe is rarely isolated: dissipation, decoherence, and relaxation typically wash out the modes that carry parameter information, limiting the attainable QFI and imposing tradeoffs among sensitivity, speed, and control complexity~\cite{gaikwad2024entanglement,saleem2023optimal,alipour2014quantum,demkowicz2014using,huang2018noise,gorecki2025interplay,zhai2023control,gebbia2020two,albarelli2023invasiveness,aiache2025nonequilibrium,brenes2023multispin,mann2025quantum}. 
Related developments in many-body, critical, nonunitary, and entanglement-enabled sensing also show that engineered open dynamics can strongly modify metrological performance~\cite{montenegro2025quantum,hotter2024combining,xiao2026observation,huang2024entanglement,alushi2024optimality,beaulieu2025criticality,sarkar2025exponentially,long2022entanglement,rahim2024entanglement,wang2018entanglement,zou2025enhancing,alushi2025collective}.

Metrology in open quantum systems encompasses both steady-state and transient metrology. Steady-state metrology is robust against timing errors after relaxation~\cite{plodzien2018few,correa2015individual}, but their sensitivity is bounded by the steady-state susceptibility and they cannot encode the elapsed evolution time once steady-state is reached. By contrast, transient metrology can reshape nonequilibrium trajectories before readout~\cite{xie2026direct,montenegro2025quantum,cavina2018bridging,kuffer2025sensing,rosenberg2026witnessing}, and these transient dynamics can effectively improve temperature estimation~\cite{aiache2024non,glatthard2022bending,gebbia2020two,sekatski2022optimal,xu2023non,frazao2024coherence,aiache2025nonequilibrium,ravell2024strongly,saha2026precision}. In particular, recent work has shown that anomalous transient relaxation effects of the Mpemba type can enhance thermometry~\cite{chattopadhyay2026anomaly,li2026quantum}.  However, these advantages demand precise control over state initialization, temporal dynamics, and measurement readout. 
Thus open quantum system metrology faces a practical tradeoff:  the fully relaxed equilibrium mechanism can suffer from prohibitively long thermalization times, whereas the precise dynamical control required for nonequilibrium protocols is challenging to implement and calibrate with high fidelity.
  
 A reset channel reinjects the probe into a prescribed target state at a tunable rate~\cite{rose2018spectral,mukherjee2018quantum,perfetto2022thermodynamics}.
 Previous studies have demonstrated that reset protocols can accelerate quantum relaxation by reshaping the system's Liouvillian eigenmodes~\cite{bao2025accelerating,solanki2025universal}. Motivated by this modal-control perspective,  we show that a reset channel can enhance both the steady-state and transient metrology. We derive a
 unified QFI response identity with respect to the reset
 rate and a strict first-order sufficient condition for local precision enhancement. 
 Reset improves steady-state thermometry only inside an exact temperature window; shifts and broadens finite-time high-QFI windows for transient thermometry; produces immediate short-time
 enhancement for time estimation through reset-induced drift; and boosts coherence-based frequency metrology even when energy-basis population readout carries no frequency information.
 Interestingly, a general short-time analysis further shows a sharp transient distinction: for static-parameter estimation in diagonal dynamics, reset does not enhance the leading-order QFI in the short-time limit; any advantage enters only at subleading order in
 time. For time estimation, in contrast, the reset contribution enters the drift immediately, and the leading contribution can scale quadratically with the reset rate.
 Thus reset provides a minimal nonequilibrium control resource for dissipative quantum sensing.

\paragraph*{Setup.---}
We consider a finite-dimensional open quantum probe governed by a parameter-dependent Lindbladian \(\mathcal L_\theta\)~\cite{lindblad1976generators} 
\begin{equation}
	\dot\rho
	=
	\mathcal L_\theta(\rho)
	=
	-i[H_\theta,\rho]
	+
	\sum_\mu
	\left[
	J_{\mu,\theta}\rho J_{\mu,\theta}^\dagger
	-
	\frac12
	\{J_{\mu,\theta}^\dagger J_{\mu,\theta},\rho\}
	\right],
	\label{eq:lindblad}
\end{equation}
where \(\theta\) is the parameter to be estimated. The parameter may enter  the Hamiltonian, the jump operators, or the rates absorbed into \(J_{\mu,\theta}\).
Without reset, an initial state \(\rho_0\) evolves as
\begin{equation}
\rho_{\theta,0}(t)	=e^{t\mathcal L_\theta}\rho_0 .
\label{eq:unreset}
\end{equation}

Assuming that \(\mathcal L_\theta\) is diagonalizable and  has a unique relaxing (globally attractive) steady state, we introduce a biorthogonal mode decomposition~\cite{campaioli2024quantum}
\(\mathcal L_\theta(R_j)=\lambda_jR_j\),
\(\mathcal L_\theta^\dagger(L_j)=\lambda_j^*L_j\), and
\(\Tr(L_j^\dagger R_k)=\delta_{jk}\).
The eigenvalues are ordered as
$0 \ge \operatorname{Re}(\lambda_j) \ge \operatorname{Re}(\lambda_{j+1})$ for \(j \ge 1\),with \(\operatorname{Re}\lambda_j<0\) for \(j\geq2\),
and the corresponding right zero mode is the unreset steady state,
$R_1=\rho_\theta^{\rm ss}.$
 The modal amplitudes are
$c_j(\theta)=\Tr[ L_j^\dagger(\theta)\rho_0],$
so that the unreset trajectory admits the expansion
\begin{equation}
	\rho_{\theta,0}(t)=\rho_\theta^{\rm ss}+\sum_{j\ge2}
    c_j(\theta)e^{\lambda_j(\theta)t}R_j(\theta).
	\label{eq:unresetmodal}
\end{equation}

The ultimate precision is bounded by the quantum Cramér--Rao bound~\cite{paris2009quantum,mukhopadhyay2025current}:
$\mathrm{Var}(\hat{\theta})\ge \frac{1}{M F[\rho_\theta(t)]}$,
where \(\mathrm{Var}(\hat{\theta})\) is the
variance of any unbiased estimator of \(\theta\), \(M\) is the number of repetitions, and $F[\rho_\theta(t)]$ is the QFI:
\begin{subequations}
	\label{eq:qfi}
	\begin{align}
		F[\rho_{\theta,0}(t)]
		&=
		\mathrm{Tr}\left[\rho_{\theta,0}(t)\mathbb{L}_{\theta}^2(t)\right],
		\label{eq:qfi-a}
		\\
		\partial_\theta\rho_{\theta,0}(t)
		&=
		\frac12
		\left\{\rho_{\theta,0}(t),\mathbb{L}_{\theta}(t)\right\}.
		\label{eq:qfi-b}
	\end{align}
\end{subequations}
Here \(\mathbb{L}_{\theta}(t)\) is the symmetric logarithmic derivative (SLD), which is defined by Eq.~\eqref{eq:qfi-b}.

\paragraph*{Reset-enhanced steady-state metrology.---}
We now add a reset channel that reinjects the probe into a fixed target state \(\rho_\delta\) at rate \(r\)~\cite{rose2018spectral,bao2025accelerating}:
\begin{equation}
	\dot\rho
	=\Lind_\theta(\rho)+
	r\bigl(\rho_\delta\Tr\rho-\rho\bigr)
	\equiv
	\Lind_{\theta,r}(\rho).
	\label{eq:master}
\end{equation}
For normalized states this reduces to
\(\dot\rho=\Lind_\theta(\rho)+r(\rho_\delta-\rho)\).
Using the biorthogonal modes of the unreset Lindbladian introduced above, define the reset-target overlaps
\(d_j(\theta)=\Tr[ L_j^\dagger(\theta)\rho_\delta].\)
The reset-modified steady state is then exactly
\begin{equation}
	\rho_{\theta,r}^{\mathrm{ss}}
	=\lim_{t\to\infty}\rho_{\theta,r}(t)
	=\rho_\theta^{\mathrm{ss}}-\sum_{j\ge2}\frac{r\,d_j(\theta)}
	{\lambda_j(\theta)-r}R_j(\theta).
	\label{eq:ssmodal}
\end{equation}
Equation~\eqref{eq:ssmodal} shows that continuous reset admixes nonsteady Liouvillian modes into the steady probe. 
The modal weights are controlled by the target overlaps \(d_j\) and by the resolvent factor \((\lambda_j-r)^{-1}\), so modes close to the origin are preferentially weighted.

For any \(C^2\), full-rank reset-dependent state \(\varrho_{\theta,r}\), a direct variation of the QFI gives the exact response identity (see End Matter):
\begin{equation}
	\partial_r F[\varrho_{\theta,r}]
	= 2\Tr\!\left[(\partial_\theta\partial_r\varrho_{\theta,r})\mathbb{L}_{\theta,r}\right]
- \Tr\!\left[(\partial_r\varrho_{\theta,r})\mathbb{L}_{\theta,r}^2\right],
	\label{eq:qfivar}
\end{equation}
where \(\mathbb{L}_{\theta,r}\) is the SLD of \(\varrho_{\theta,r}\). 
This identity applies both to the steady state \(\varrho_{\theta,r}=\rho_{\theta,r}^{\mathrm{ss}}\)
and to the transient state \(\varrho_{\theta,r}=\rho_{\theta,r}(t)\)
at fixed readout time \(t\). 
It is not itself an advantage condition; finite-\(r\) advantage is determined by the integrated response from \(0\) to \(r\).
Expand Eq.~\eqref{eq:ssmodal} to first order in \(r\), \(\rho_{\theta,r}^{\rm ss} = \rho_\theta^{\rm ss} + r\rho_{1,\theta}^{\rm ss} +O(r^2), \) 
\(\rho_{1,\theta}^{\rm ss} = \left.\partial_r\rho_{\theta,r}^{\rm ss}\right|_{r=0} = -\sum_{j\ge2}({d_j(\theta)R_j(\theta)})/{\lambda_j(\theta)}.\)
Thus a strict first-order sufficient condition for local steady-state enhancement is 

\begin{equation}
	\left.\partial_r F[\rho_{\theta,r}^{\mathrm{ss}}]\right|_{r=0}
	=2\Tr\!\left[(\partial_\theta\rho_{1,\theta}^{\mathrm{ss}})\mathbb{L}_\theta^{(0)}\right]- \Tr\!\left[\rho_{1,\theta}^{\mathrm{ss}}(\mathbb{L}_\theta^{(0)})^2\right]>0
	\label{eq:smallrcriterion}
\end{equation}
where \(\mathbb{L}_\theta^{(0)}\) is the SLD of the unreset steady state \(\rho_\theta^{\mathrm{ss}}\). 
Equation~\eqref{eq:smallrcriterion} states that reset is locally beneficial when the reset-induced  steady correction has  a large positive sensitivity to the estimated parameter.

\paragraph*{Reset-enhanced transient metrology.---}
For a reset pulse applied during \(0\le t\le t_s\), the reset-modified transient state is denoted by \(\rho_{\theta,r}(t)\). With the initial-state and reset-target modal overlaps \(c_k(\theta)=\Tr[ L_k^\dagger(\theta)\rho_0],d_k(\theta)=\Tr[ L_k^\dagger(\theta)\rho_\delta],\)
the exact state during the pulse can be written, suppressing the \(\theta\)-dependence for compactness, as~\cite{bao2025accelerating}
\begin{subequations}
	\label{eq:transientmodal}
	\begin{align}
		\rho_{\theta,r}(t)
		&=
		\rho_\theta^{\mathrm{ss}}
		+
		\sum_{k\ge2}
		c_k^{\,r}(t)e^{\lambda_k t}R_k,
		\label{eq:transientmodal-a}
		\\
		c_k^{\,r}(t)
		&=
		\left(
		c_k-\frac{r d_k}{r-\lambda_k}
		\right)e^{-rt}
		+
		\frac{r d_k}{r-\lambda_k}e^{-\lambda_k t},
		\label{eq:transientmodal-b}
	\end{align}
\end{subequations}
after the pulse is switched off, the amplitude multiplying \(e^{\lambda_k t}\) is fixed at \(c_k^{\,r}(t_s)\), and the subsequent evolution proceeds under the original spectrum.

At any fixed readout time \(t\), write \(\rho_{\theta,r}(t) =
\rho_{\theta,0}(t) + r\left.\partial_r\rho_{\theta,r}(t)\right|_{r=0} +O(r^2)\). Equation~\eqref{eq:qfivar} then gives \(	\left.\partial_rF[\rho_r(t;\theta)]\right|_{r=0}=2\Tr[(\partial_\theta\left.\partial_r\rho_{\theta,r}(t)\right|_{r=0})\mathbb{L}_t^{(0)}]
-\Tr[\left.\partial_r\rho_{\theta,r}(t)\right|_{r=0}(\mathbb{L}_t^{(0)})^2]>0.\) A positive value is a strict first-order sufficient condition for local transient enhancement.

A universal short-time distinction appears when the dynamics remains diagonal in a parameter-independent basis. Assuming full support of the initial population, \((\vp_0)_i>0\) for all components entering the Fisher-information sum, the QFI reduces to the classical Fisher information~\cite{paris2009quantum}
\begin{equation}
	F_\theta^{(r)}(t)=\sum_i
	\frac{[\partial_\theta p_{\theta,r,i}(t)]^2}
	{p_{\theta,r,i}(t)} .
	\label{eq:diagonal-fi}
\end{equation}

Let \(\vp_{\theta,r}(t)\) denote the population vector, and let \(W_\theta\) be the classical rate generator induced by \(\mathcal L_\theta\) in this fixed basis.
With initial population \(\vp_0\) and reset-target population \(\vp_\delta\), the populations obey \(\dot{\vp}_{\theta,r}=W_\theta\vp_{\theta,r}+r(\vp_\delta-\vp_{\theta,r}),\)
and hence \(\vp_{\theta,r}(t)=\vp_0+t\bigl[W_\theta\vp_0+r(\vp_\delta-\vp_0)\bigr]+O(t^2).\)
For a static parameter \(\theta\), with \(\vp_0\) and \(\vp_\delta\) independent of \(\theta\), and under the same full-support assumption,
\begin{equation}
	F_\theta^{(r)}(t)
	=t^2\sum_i
	\frac{\bigl[(\partial_\theta W_\theta)\vp_0\bigr]_i^2}
	{(\vp_0)_i}+O(t^3),
	\label{eq:shorttime-static}
\end{equation}
so reset does not modify the leading \(t^2\) term.
By contrast, the instantaneous time-estimation Fisher information is
\begin{equation}
	F_t^{(r)}(0)
	=
	\sum_i
	\frac{\bigl[W_\theta\vp_0+r(\vp_\delta-\vp_0)\bigr]_i^2}{(\vp_0)_i},
	\label{eq:shorttime-time}
\end{equation}
which is reset dependent already at \(t=0\).
Thus static-parameter sensing gains require finite information-accumulation time, whereas time estimation can benefit immediately from the reset-induced drift.

\begin{figure}[t]
	\centering
	\includegraphics[width=0.50\textwidth]{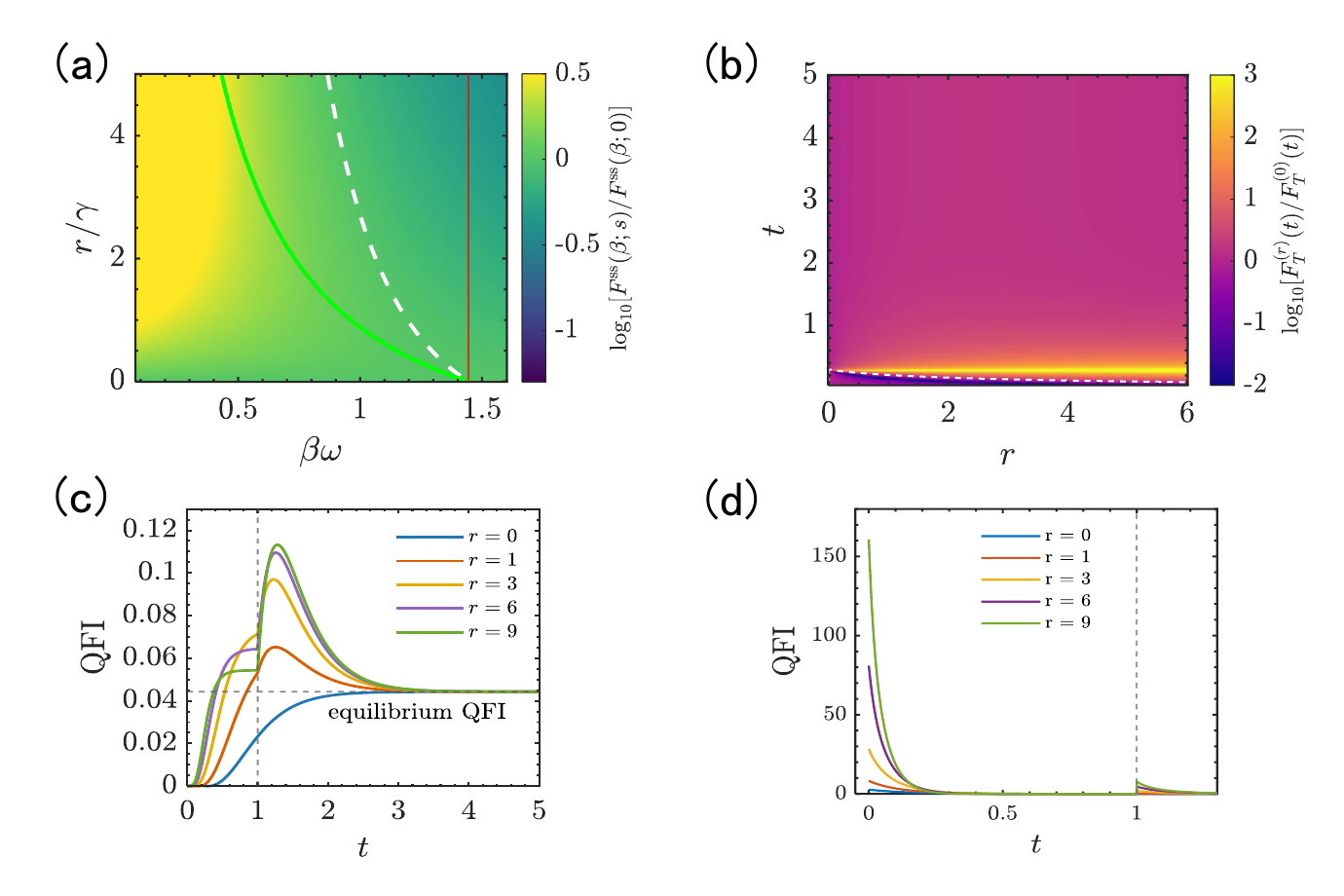}
\caption{
	Analytic thermal-qubit case studies for reset-enhanced thermometry and time metrology.
	(a) Steady-state thermometric gain 
	\(\log_{10}[F^{\mathrm{ss}}(\beta;s)/F^{\mathrm{ss}}(\beta;0)]\).
	The green solid curve gives the optimal reset rate \(r_{\rm opt}/\gamma\), the white dashed curve marks the unit-gain boundary, and the red vertical line indicates the theoretical boundary \(\beta\omega=\ln(2+\sqrt{5})\).
	(b) Transient thermometric gain \(\log_{10}[F_T^{(r)}(t)/F_T^{(0)}(t)]\). The white dashed curve marks the unit-gain boundary.
	(c) Transient thermometry QFI. The horizontal gray dashed line marks the equilibrium QFI.
	(d) Transient time metrology. 
	The vertical dashed lines in (c,d) mark reset switch-off at \(t_s=1\).
	In (b)--(d), \(\omega=\gamma=1\), \(T=1.5\); in (a)--(d) \(p_\delta=0\).
}
	\label{fig:2}
\end{figure}

\paragraph*{Steady-state thermometry: exact advantage window.---}
As a solvable benchmark, we consider a thermal qubit with
\(H=(\omega/2)\sigma_z\), 
\(\sigma_z=|1\rangle\langle1|-|0\rangle\langle0|\), and
$\mathcal L_\beta(\rho) = -i\left[\frac{\omega}{2}\sigma_z, \rho\right] + \gamma_\downarrow\mathcal{D}[\sigma_-]\rho+\gamma_\uparrow\mathcal{D}[\sigma_+]\rho$, $\mathcal D[A]\rho= A\rho A^\dagger - \frac12\{A^\dagger A,\rho\}. $
Here \(\sigma_-=|0\rangle\langle1|\), \(\sigma_+=|1\rangle\langle0|\), and the estimated parameter is
\(\theta=\beta=(k_B T)^{-1}\) with \(k_{\rm B}=1\). 
The bosonic transition rates are
$\gamma_\uparrow=\gamma n(\beta)$, $\gamma_\downarrow=\gamma[n(\beta)+1]$, where \(n(\beta)=(e^{\beta\omega}-1)^{-1}\).
We reset to the ground state \(|0\rangle=(0,1)^{\rm T}\), so \(p_\delta=0\). 
Defining \(p_{\beta,r}(t)=\langle1|\rho_{\beta,r}(t)|1\rangle\), the closed excited-state population dynamics is
\begin{equation}
	\dot p_{\beta,r}
	=-(\gamma_\downarrow+\gamma_\uparrow+r)p_{\beta,r}+\gamma_\uparrow
	+rp_\delta .
	\label{eq:rateequation}
\end{equation}
 
  Setting $\hbar=1$, the reset steady-state excited-state population is
 \begin{equation}
	p_{\beta,r}^{\mathrm{ss}}=\frac{\gamma}
	{\gamma\left(e^{\beta\omega}+1\right)
		+r\left(e^{\beta\omega}-1\right)} .
	\label{eq:excited}
\end{equation}

The QFI is the classical Fisher information of the binary distribution
\(\{p_{\beta,r}^{\mathrm{ss}},1-p_{\beta,r}^{\mathrm{ss}}\}\), $F^{\mathrm{ss}}(\beta;r) = \left(\partial_\beta p_{\beta,r}^{\mathrm{ss}}\right)^2/ p_{\beta,r}^{\mathrm{ss}}\left(1-p_{\beta,r}^{\mathrm{ss}}\right)$. Writing \(q=e^{\beta\omega}\) and \(s=r/\gamma\), we obtain the exact closed form
\begin{equation}
	F^{\mathrm{ss}}(\beta;s)=\omega^2
	\frac{q^2(1+s)^2}{\bigl[(1+s)q-s\bigr]\bigl[(1+s)q-s+1\bigr]^2}.
	\label{eq:ssqfi}
\end{equation}
This gives a complete analytic characterization of the
steady-state reset advantage. The inequality
$F^{\mathrm{ss}}(\beta;s)>F^{\mathrm{ss}}(\beta;0)$ holds iff
$0<\beta\omega<\ln(2+\sqrt5)$ and $0<s<s_+(q)$, where $s_+(q)$ is the
positive root of the corresponding quadratic inequality. The optimal reset rate is $r_{\mathrm{opt}}=\gamma\max\!\left[0,(2+\sqrt5-e^{\beta\omega})/(e^{\beta\omega}-1)\right]$.
For every finite \(\beta\omega>0\), this optimum is finite.
Its divergence as \(\beta\omega\to0\) accompanies the divergence of the ideal bosonic thermalization rate \(\Gamma=\gamma\coth(\beta\omega/2)\); it is therefore a nonuniform boundary of the effective bath model rather than a prescription to realize an infinite rate. Along the optimal curve, \(p_{\beta,r_{\rm opt}}^{\rm ss}\to1/(3+\sqrt5)\), so this joint high-temperature limit does not lose rank.
The explicit form of $s_+(q)$ and the derivation of
$r_{\mathrm{opt}}$ are given in the Supplemental Material. 
   
Figure~\ref{fig:2}(a) shows that reset improves steady-state thermometry only in the high-temperature window where the ground-state reset increases the thermometric susceptibility rather than washing it out. 
In this solvable example, reset does not merely accelerate relaxation; it reshapes the steady response curve itself.

 \paragraph*{Transient thermometry and elapsed-time estimation.---}
 Throughout, thermometry and elapsed-time estimation are distinct single-parameter tasks.
 In thermometry the readout time is controlled and the temperature is unknown, whereas in time estimation the temperature and all generator parameters are calibrated and the unknown parameter is the elapsed time since initialization.
 The latter is the standard operational task of a quantum clock~\cite{buvzek1999optimal,faist2023time,prech2025optimal}; no joint estimation of temperature and time is assumed, and the QFI quantifies local precision within a calibrated operating window.
 
 Using the same thermal qubit, we now compare finite-time thermometry and time estimation. Under the reparametrization $\beta = 1/T$,  the QFI transforms as $F_T = F_\beta / T^4$. In the absence of reset, define \(\bar n(T)=[e^{\omega/T}-1]^{-1}\); the equilibrium excited-state population and the relaxation rate are
 \(p_{\rm eq}(T)=\bar n(T)/[2\bar n(T)+1]\) and \(\Gamma(T)=\gamma[2\bar n(T)+1]\). 
 The population obeys $\dot p_{T,r}=-(\Gamma+r)p_{T,r}+\Gamma p_{\rm eq}+r p_\delta$ with the reset steady value $p_{T,r}^*=(\Gamma p_{\rm eq}+r p_\delta)/(\Gamma+r)$.
 Thus, during a reset pulse \(0\le t\le t_s\),
 \begin{equation}
 	p_{T,r}(t)
 	=p_{T,r}^*+\bigl(p_0-p_{T,r}^*\bigr)e^{-(\Gamma+r)t}.
 	\label{eq:prtransient}
 \end{equation}
 Since the state remains diagonal in the energy basis, the QFI reduces to the classical Fisher information of the binary distribution
 \(\{p_{T,r}(t),1-p_{T,r}(t)\}\). Equation~\eqref{eq:diagonal-fi} gives the finite-time Fisher information for thermometry and time estimation:
 \begin{subequations}
 	\begin{align}
 		F_T^{(r)}(t)
 		&=
 		\frac{[\partial_Tp_{T,r}(t)]^2}
 		{p_{T,r}(t)[1-p_{T,r}(t)]},
 		\label{eq:FT_r}\\
 		F_t^{(r)}(t)
 		&=
 		\frac{[\partial_tp_{T,r}(t)]^2}
 		{p_{T,r}(t)[1-p_{T,r}(t)]};
 		\label{eq:Ft_r}
 	\end{align}
 \end{subequations}
 after the pulse is switched off, the dynamics reverts to the original thermalization process,
 \begin{equation}
 	p_{T,r}(t)
 	=
 	p_{\rm eq}
 	+
 	\bigl[p_{T,r}(t_s)-p_{\rm eq}\bigr]e^{-\Gamma(t-t_s)},
 	\qquad t>t_s .
 	\label{eq:prtransientoff}
 \end{equation}
 
In agreement with Eq.~\eqref{eq:shorttime-static}, thermometric gains appear only after a finite information-accumulation time. 
Large gain ratios near small unreset baselines should be interpreted together with the absolute QFI curves. 
Figure~\ref{fig:2}(b) shows the corresponding transient gain when reset is kept on until readout, \(t_s=t\), while Fig.~\ref{fig:2}(c) shows that a temporary reset pulse can move the probe into a high-\(F_T\) window earlier and, for suitable \(r\), overshoot the equilibrium thermometric Fisher information over a finite interval. 
However, transient thermometric reset should not be interpreted as a universal peak-enhancement mechanism; the global instantaneous maximum may still occur without reset for some initial conditions, as discussed in the Supplemental Material. 
Rather, for a chosen readout time or a finite readout window, an intermediate reset rate can shift and broaden the high-QFI region, improving timing robustness, whereas too strong reset pins the probe close to the parameter-blind target and suppresses temperature susceptibility.

 Time estimation behaves differently. 
 Because Eq.~\eqref{eq:shorttime-time} contains the reset-induced drift already at \(t=0\), \(F_t^{(r)}\) responds immediately to reset. 
 In the short-time limit, the reset drift contributes linearly to \(\partial_t p_{T,r}\), and hence the time-estimation Fisher information can scale quadratically with \(r\) for fixed initial and reset-target populations while \(rt\ll1\). 
 This contrast is visible in Fig.~\ref{fig:2}(d): \(F_t^{(r)}\) is sharply enhanced immediately after the reset pulse is turned on, but it rapidly decays as the trajectory approaches the reset-modified fixed point. 
 When the pulse is switched off at \(t_s=1\), the population remains continuous, whereas its velocity changes because the generator is abruptly changed back to the original thermal dynamics. 
 For the parameters used here, the short-time enhancement relative to the unreset trajectory can exceed two orders of magnitude. 
 The explicit expressions of Eqs.~\eqref{eq:FT_r} and \eqref{eq:Ft_r}, the small-reset response coefficients, and the dependence on the initial excited-state population are given in the Supplemental Material.
 
 \begin{figure}[t]
 	\centering
 	\includegraphics[width=0.50\textwidth]{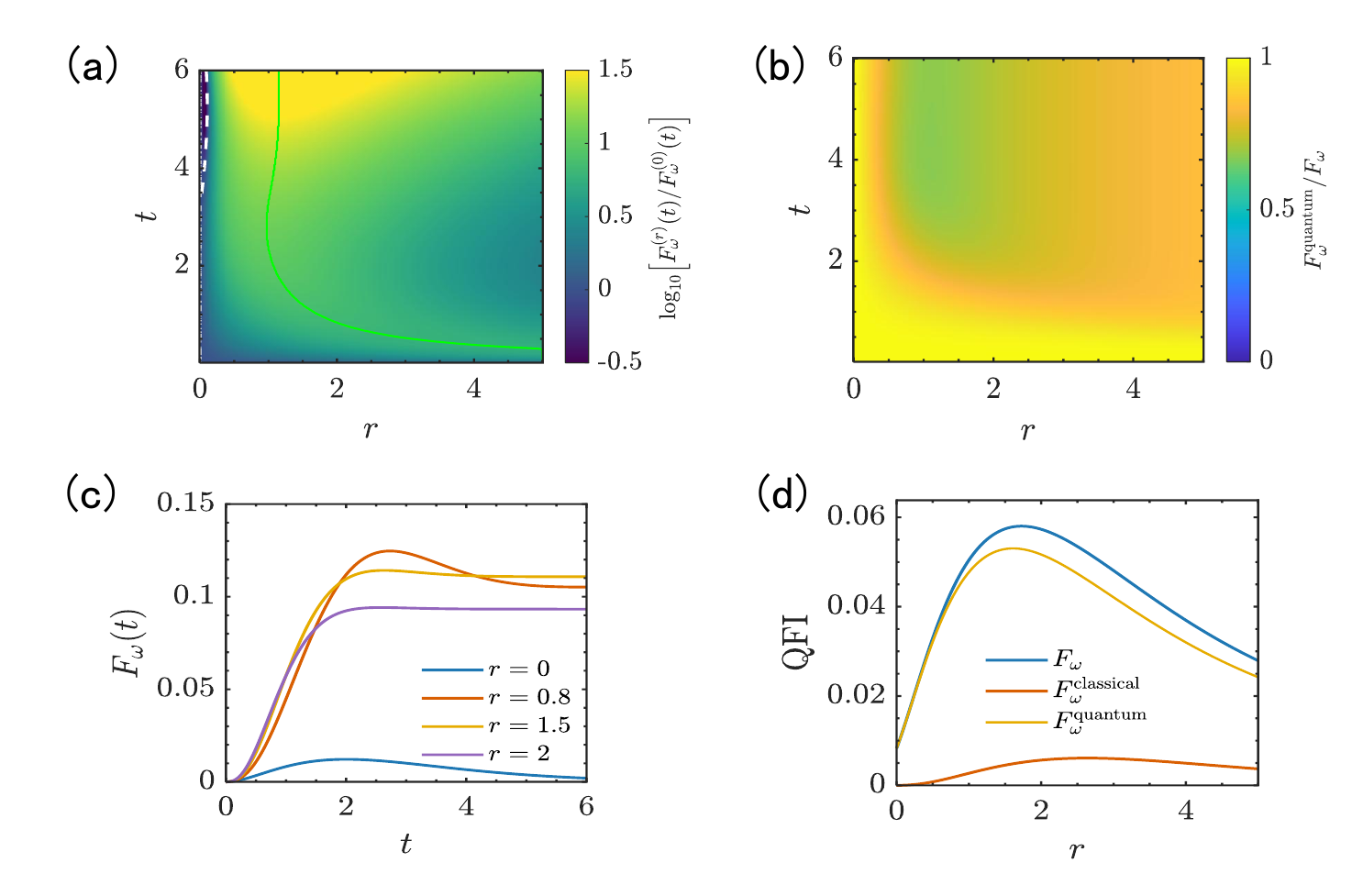}
 	\caption{
 		Coherent qubit frequency metrology.
 		(a) Frequency-metrology gain map $\log_{10}[F_\omega^{(r)}(t)/F_\omega^{(0)}(t)]$.
 		The white contour marks unit gain and the green solid curve is the optimal reset rate $r$.
 		(b) Quantum fraction $F_\omega^{\rm quantum}/F_\omega$.
 		(c) Line cuts at representative reset rates.
 		(d) Bloch-vector decomposition $F_\omega=F_\omega^{\mathrm{classical}}+F_\omega^{\mathrm{quantum}}$ at fixed $t=1$. At $t=1$ the optimum is $r\approx1.73$, with $F_\omega^{(r)}/F_\omega^{(0)}\approx6.96$ and $F_\omega^{(r)\mathrm{quantum}}/F_\omega^{(r)}\approx0.91$.
 		Here``classical'' and ``quantum'' denote the density-matrix eigenvalue and eigenvector contributions, respectively, not the two explicit terms in Eq.~\eqref{eq:freqfi}.
 		Parameters: $\gamma=\omega=k_B=\hbar=1$ and $\rho_0=\begin{psmallmatrix}0.9&0.075\\0.075&0.1\end{psmallmatrix}$. 
 	}
 	\label{fig:3}
 \end{figure}
 
 \paragraph*{Transient frequency metrology: quantum gain.---}
 To expose a purely coherence-borne advantage, we next consider a qubit with amplitude damping, where
 $\mathcal L_\omega(\rho) = -i\left[\frac{\omega}{2}\sigma_z, \rho\right] + \gamma\mathcal{D}[\sigma_-]\rho$,
 and reset the system to $\rho_\delta = |+\rangle\langle +|$, where $|+\rangle = \left(|0\rangle + |1\rangle\right)/\sqrt{2}$. Writing
 $\rho_{\omega,r}(t)=\begin{pmatrix}
 	p_{\omega,r}(t) & c_{\omega,r}^*(t)\\
 	c_{\omega,r}(t) & 1-p_{\omega,r}(t)\end{pmatrix}$,
 one finds
 \begin{subequations}\label{eq:coherent}
 	\begin{align}
 		p_{\omega,r}(t)&=\frac{r}{2(\gamma+r)}+\left[p_0-\frac{r}{2(\gamma+r)}\right]e^{-(\gamma+r)t},\\
 		c_{\omega,r}(t)&=\frac{r/2}{\gamma/2+r-i\omega}+\left[c_0-\frac{r/2}{\gamma/2+r-i\omega}\right]e^{(i\omega-\gamma/2-r)t}.
 	\end{align}
 \end{subequations}
 
 Since $\partial_\omega p_{\omega,r}(t)=0$, energy-population measurements carry no information about \(\omega\). 
 All frequency information is encoded in the coherence \(c_{\omega,r}(t)\). 
 For the full-rank qubit trajectories considered here, the exact QFI is~\cite{zhong2013fisher} \(F_\theta=\operatorname{Tr}(\rho \mathbb{L}_\theta^2)
 =|\partial_\theta\bm s|^2	+
 \frac{(\bm s\cdot\partial_\theta\bm s)^2}{1-|\bm s|^2}\), where \(\bm s\) is the Bloch vector of the qubit state. Therefore:
 \begin{equation}
 	F_\omega^{(r)}(t)=4|\partial_\omega c_{\omega,r}|^2+\frac{\left[\partial_\omega|c_{\omega,r}|^2\right]^2}{p_{\omega,r}(1-p_{\omega,r})-|c_{\omega,r}|^2}.
 	\label{eq:freqfi}
 \end{equation}
 
The two explicit terms in Eq.~\eqref{eq:freqfi} should not be confused with the spectral decomposition used in Fig.~\ref{fig:3} (b) and (d).
Writing \(s=|\boldsymbol{s}|\) and \(\hat{\boldsymbol{s}}=\boldsymbol{s}/s\), we obtain
\(F_\omega^{\rm classical}=(\partial_\omega s)^2/(1-s^2)\) and
\(F_\omega^{\rm quantum}=s^2|\partial_\omega\hat{\boldsymbol{s}}|^2\).
Although \(\partial_\omega p_{\omega,r}=0\), \(p_{\omega,r}\) and \(1-p_{\omega,r}\) are energy-basis populations rather than density-matrix eigenvalues; the eigenvalues \((1\pm s)/2\) generally inherit \(\omega\) dependence from \(|c_{\omega,r}|\) for \(r>0\).
For \(r=0\), this reduces to
\(F_\omega^{(0)}(t)=4|c_0|^2t^2e^{-\gamma t}\).
The detailed derivation of Eq.~\eqref{eq:freqfi}, the classical-quantum decomposition of the Bloch vector, and the verification of the positive first-order reset response for the plotted parameters obtained from Eq.~\eqref{eq:qfivar}, are provided in the Supplemental Material.
 
 Reset enhances frequency estimation by replenishing the coherent Liouvillian modes that carry the \(\omega\)-dependent phase.  Figure~\ref{fig:3}(a) shows a broad frequency-metrology gain region, while Fig.~\ref{fig:3}(c) confirms that the enhancement persists over finite readout times for representative reset rates. The quantum-fraction map in Fig.~\ref{fig:3}(b), together with the Bloch decomposition in Fig.~\ref{fig:3}(d), shows that the optimized gain is dominated by the eigenvector contribution \(F_\omega^{\rm quantum}\), rather than by the eigenvalue contribution \(F_\omega^{\rm classical}\). Reset-enhanced frequency estimation is therefore genuinely coherence-driven. Since the population dynamics is independent of \(\omega\) in this model, the advantage cannot be accessed by energy-population readout alone; it requires phase-sensitive measurements. This example cleanly separates coherence engineering from population engineering and shows that reset can enhance metrology even when the information is encoded only in quantum coherences.

\paragraph*{Conclusion and outlook.---}
We have shown that reset can act as a metrological resource in open quantum systems by converting dissipative modal control into QFI control. The reset-rate response identity and the small-reset sufficient condition provide a general diagnostic for local enhancement in both steady-state and transient regimes.  Our short-time analysis further reveals a fundamental transient distinction: static-parameter sensing has no leading short-time reset gain in fixed-basis diagonal dynamics, whereas time estimation can benefit immediately from the reset-induced drift, with a contribution scaling quadratically in the reset rate for \(rt\ll1\). In transient frequency estimation, we find that the reset-enhanced QFI is dominated by the genuinely quantum eigenvector contribution rather than by the density-matrix-eigenvalue contribution. These results show reset as a minimal dissipative control resource for enhancing metrological precision. Future directions include multilevel and many-body probes, adaptive reset schedules, constrained measurements, and multiparameter estimation~\cite{szczykulska2016multi,demkowicz2020multi,mihailescu2026metrological,brady2026precision}.

\paragraph*{Acknowledgments.---}
 We thank Siyuan Bai for valuable discussions. This work is partly supported
 by National Natural Science Foundation of China (Grant No. 12374178), National Key Research and Development Program of Ministry of Science and Technology (Grants
 No. 2022YFA1204700 and No. 2021YFA1200503). R. B. was supported by JSPS KAKENHI Grant No. 25KJ0766.

\section*{End Matter}

\textit{Derivation of the Fisher-information response identity.}---%
Here we derive Eq.~\eqref{eq:qfivar}. Let \(\varrho_{\theta,r}\) be a smooth reset-dependent state  and let \(L_{\theta,r}\) be its symmetric logarithmic derivative (SLD), defined by
\begin{equation}
	\partial_\theta \varrho_{\theta,r}
	=
	\frac12\left\{\varrho_{\theta,r},L_{\theta,r}\right\}
	\equiv
	\mathcal J_{\varrho_{\theta,r}}(L_{\theta,r}),
	\label{eq:em_sld}
\end{equation}
where \(\mathcal J_{\rho}(X)=\frac12(\rho X+X\rho)\). The Fisher information is
\begin{equation}
	F[\varrho_{\theta,r}]
	=
	\Tr\!\left[(\partial_\theta\varrho_{\theta,r})L_{\theta,r}\right]
	=
	\Tr\!\left[\varrho_{\theta,r}L_{\theta,r}^2\right].
	\label{eq:em_fi}
\end{equation}
Taking a derivative of the first expression in Eq.~\eqref{eq:em_fi} with respect to \(r\) gives
\begin{equation}
	\partial_r F[\varrho_{\theta,r}]
	=
	\Tr\!\left[(\partial_r\partial_\theta\varrho_{\theta,r})L_{\theta,r}\right]
	+
	\Tr\!\left[(\partial_\theta\varrho_{\theta,r})(\partial_r L_{\theta,r})\right].
	\label{eq:em_first}
\end{equation}
Using Eq.~\eqref{eq:em_sld}, the second term becomes
\begin{equation}
	\Tr\!\left[(\partial_\theta\varrho_{\theta,r})(\partial_r L_{\theta,r})\right]
	=
	\Tr\!\left[\mathcal J_{\varrho_{\theta,r}}(L_{\theta,r})(\partial_r L_{\theta,r})\right].
\end{equation}
Since the Jordan superoperator is self-adjoint under the Hilbert--Schmidt inner product, \(\Tr\!\left[\mathcal J_{\rho}(A)B\right]=\Tr\!\left[A\mathcal J_{\rho}(B)\right]\), we obtain
\begin{equation}
	\Tr\!\left[(\partial_\theta\varrho_{\theta,r})(\partial_r L_{\theta,r})\right]
	=
	\Tr\!\left[L_{\theta,r}\,\mathcal J_{\varrho_{\theta,r}}(\partial_r L_{\theta,r})\right].
	\label{eq:em_second}
\end{equation}
Differentiating the SLD equation \(\partial_\theta\varrho_{\theta,r}=\mathcal J_{\varrho_{\theta,r}}(L_{\theta,r})\) with respect to \(r\) yields
\begin{equation}
	\partial_r\partial_\theta\varrho_{\theta,r}
	=
	\mathcal J_{\partial_r\varrho_{\theta,r}}(L_{\theta,r})
	+
	\mathcal J_{\varrho_{\theta,r}}(\partial_r L_{\theta,r}),
	\label{eq:em_sldvar}
\end{equation}
where \(\mathcal J_{\partial_r\varrho_{\theta,r}}(L_{\theta,r})=\frac12\left\{\partial_r\varrho_{\theta,r},L_{\theta,r}\right\}\). Rearranging gives
\begin{equation}
	\mathcal J_{\varrho_{\theta,r}}(\partial_r L_{\theta,r})
	=
	\partial_r\partial_\theta\varrho_{\theta,r}
	-
	\frac12\left\{\partial_r\varrho_{\theta,r},L_{\theta,r}\right\}.
	\label{eq:em_jinv}
\end{equation}
Substituting Eq.~\eqref{eq:em_jinv} into Eq.~\eqref{eq:em_second} and using trace cyclicity:
\begin{align}
	\Tr\!\left[(\partial_\theta\varrho_{\theta,r})(\partial_r L_{\theta,r})\right]
	&=
	\Tr\!\left[(\partial_r\partial_\theta\varrho_{\theta,r})L_{\theta,r}\right]
	-
	\Tr\!\left[(\partial_r\varrho_{\theta,r})L_{\theta,r}^2\right]
	\label{eq:em_fourth}
\end{align}
Finally, substituting Eq.~\eqref{eq:em_fourth} into Eq.~\eqref{eq:em_first} gives
\begin{equation}
	{
		\partial_r F[\varrho_{\theta,r}]
		=
		2\Tr\!\left[(\partial_\theta\partial_r\varrho_{\theta,r})L_{\theta,r}\right]
		-
		\Tr\!\left[(\partial_r\varrho_{\theta,r})L_{\theta,r}^2\right].
	}
	\label{eq:em_qfivar}
\end{equation}
This is Eq.~\eqref{eq:qfivar}.
It applies to both \(\varrho_{\theta,r}=\rho_{\theta,r}^{\rm ss}\) and
\(\varrho_{\theta,r}=\rho_{\theta,r}(t)\) at fixed readout time \(t\).
The derivation above is directly valid on any full-rank patch.
It extends unchanged to a \(C^2\) constant-rank family by choosing the minimal SLD \(L_{\theta,r}=\mathcal J_{\varrho_{\theta,r}}^{\dagger}(\partial_\theta\varrho_{\theta,r})\), because the Moore--Penrose inverse is then smooth.
At a rank-changing point the SLD QFI need not be differentiable~\cite{vsafranek2017discontinuities}, so Eq.~\eqref{eq:qfivar} is applied only on each fixed-rank side or through a well-defined one-sided limit.
All transient states evaluated here at finite readout time, and all thermal steady states at finite temperature and reset rate, are full rank; no derivative at \(r=\infty\) is invoked.

	% reference
\bibliography{reference}

@article{bao2025accelerating,
	title = {Accelerating Quantum Relaxation via Temporary Reset: A Mpemba-Inspired Approach},
	author = {Bao, Ruicheng and Hou, Zhonghuai},
	journal = {Phys. Rev. Lett.},
	volume = {135},
	issue = {15},
	pages = {150403},
	numpages = {10},
	year = {2025},
	month = {Oct},
	publisher = {American Physical Society},
	doi = {10.1103/g94p-7421},
	url = {https://link.aps.org/doi/10.1103/g94p-7421}
}

@article{chattopadhyay2026anomaly,
	title={Anomaly to Resource: The Mpemba Effect in Quantum Thermometry},
	author={Chattopadhyay, Pritam and Santos, Jonas F. G. and Misra, Avijit},
	journal={arXiv preprint},
	year={2026},
	eprint={2601.05046},
	archivePrefix={arXiv},
	primaryClass={quant-ph},
	version={1},
	doi={10.48550/arXiv.2601.05046}
}

@article{giovannetti2006quantum,
	title={Quantum metrology},
	author={Giovannetti, Vittorio and Lloyd, Seth and Maccone, Lorenzo},
	journal={Physical review letters},
	volume={96},
	number={1},
	pages={010401},
	year={2006},
	publisher={APS}
}

@article{toth2014quantum,
	title={Quantum metrology from a quantum information science perspective},
	author={T{\'o}th, G{\'e}za and Apellaniz, Iagoba},
	journal={Journal of Physics A: Mathematical and Theoretical},
	volume={47},
	number={42},
	pages={424006},
	year={2014},
	publisher={IOP Publishing}
}

@article{szczykulska2016multi,
	title={Multi-parameter quantum metrology},
	author={Szczykulska, Magdalena and Baumgratz, Tillmann and Datta, Animesh},
	journal={Advances in Physics: X},
	volume={1},
	number={4},
	pages={621--639},
	year={2016},
	publisher={Taylor \& Francis}
}

@article{braunstein1994statistical,
	title={Statistical distance and the geometry of quantum states},
	author={Braunstein, Samuel L and Caves, Carlton M},
	journal={Physical Review Letters},
	volume={72},
	number={22},
	pages={3439},
	year={1994},
	publisher={APS}
}

@article{paris2009quantum,
	title={Quantum estimation for quantum technology},
	author={Paris, Matteo GA},
	journal={International Journal of Quantum Information},
	volume={7},
	number={supp01},
	pages={125--137},
	year={2009},
	publisher={World Scientific}
}

@article{giovannetti2011advances,
	title={Advances in quantum metrology},
	author={Giovannetti, Vittorio and Lloyd, Seth and Maccone, Lorenzo},
	journal={Nature photonics},
	volume={5},
	number={4},
	pages={222--229},
	year={2011},
	publisher={Nature Publishing Group UK London}
}

@article{demkowicz2020multi,
	title={Multi-parameter estimation beyond quantum Fisher information},
	author={Demkowicz-Dobrza{\'n}ski, Rafa{\l} and G{\'o}recki, Wojciech and Gu{\c{t}}{\u{a}}, M{\u{a}}d{\u{a}}lin},
	journal={Journal of Physics A: Mathematical and Theoretical},
	volume={53},
	number={36},
	pages={363001},
	year={2020},
	publisher={IOP Publishing}
}

@article{liu2024fully,
	title={Fully-optimized quantum metrology: framework, tools, and applications},
	author={Liu, Qiushi and Hu, Zihao and Yuan, Haidong and Yang, Yuxiang},
	journal={Advanced Quantum Technologies},
	volume={7},
	number={12},
	pages={2400094},
	year={2024},
	publisher={Wiley Online Library}
}

@article{brady2026precision,
	title={Precision Limits of Multiparameter Markovian-Noise Metrology},
	author={Brady, Anthony J and Wang, Yu-Xin and Garc{\'\i}a-Pintos, Luis Pedro and Gorshkov, Alexey V},
	journal={arXiv preprint arXiv:2604.14298},
	year={2026}
}

@article{mihailescu2026metrological,
	title={Metrological symmetries in singular quantum multi-parameter estimation},
	author={Mihailescu, George and Sarkar, Saubhik and Bayat, Abolfazl and Campbell, Steve and Mitchell, Andrew K},
	journal={Quantum Science and Technology},
	volume={11},
	number={1},
	pages={015006},
	year={2026},
	publisher={IOP Publishing}
}

@article{polino2020photonic,
	title={Photonic quantum metrology},
	author={Polino, Emanuele and Valeri, Mauro and Spagnolo, Nicol{\`o} and Sciarrino, Fabio},
	journal={AVS Quantum Science},
	volume={2},
	number={2},
	year={2020},
	publisher={AIP Publishing}
}

@article{mitchell2004super,
	title={Super-resolving phase measurements with a multiphoton entangled state},
	author={Mitchell, Morgan W and Lundeen, Jeff S and Steinberg, Aephraem M},
	journal={Nature},
	volume={429},
	number={6988},
	pages={161--164},
	year={2004},
	publisher={Nature Publishing Group UK London}
}

@article{pezze2007phase,
	title={Phase detection at the quantum limit with multiphoton Mach-Zehnder interferometry},
	author={Pezz{\'e}, L and Smerzi, A and Khoury, G and Hodelin, JF and Bouwmeester, D},
	journal={Physical review letters},
	volume={99},
	number={22},
	pages={223602},
	year={2007},
	publisher={APS}
}

@article{gross2010nonlinear,
	title={Nonlinear atom interferometer surpasses classical precision limit},
	author={Gross, Christian and Zibold, Tilman and Nicklas, Eike and Esteve, Jerome and Oberthaler, Markus K},
	journal={Nature},
	volume={464},
	number={7292},
	pages={1165--1169},
	year={2010},
	publisher={Nature Publishing Group UK London}
}

@article{ono2013entanglement,
	title={An entanglement-enhanced microscope},
	author={Ono, Takafumi and Okamoto, Ryo and Takeuchi, Shigeki},
	journal={Nature communications},
	volume={4},
	number={1},
	pages={2426},
	year={2013},
	publisher={Nature Publishing Group UK London}
}

@article{ding2022enhanced,
	title={Enhanced metrology at the critical point of a many-body Rydberg atomic system},
	author={Ding, Dong-Sheng and Liu, Zong-Kai and Shi, Bao-Sen and Guo, Guang-Can and M{\o}lmer, Klaus and Adams, Charles S},
	journal={Nature Physics},
	volume={18},
	number={12},
	pages={1447--1452},
	year={2022},
	publisher={Nature Publishing Group UK London}
}

@article{leibfried2004toward,
	title={Toward Heisenberg-limited spectroscopy with multiparticle entangled states},
	author={Leibfried, Dietrich and Barrett, Murray D and Schaetz, Tobias and Britton, Joseph and Chiaverini, John and Itano, Wayne M and Jost, John D and Langer, Christopher and Wineland, David J},
	journal={Science},
	volume={304},
	number={5676},
	pages={1476--1478},
	year={2004},
	publisher={American Association for the Advancement of Science}
}

@article{bonato2016optimized,
	title={Optimized quantum sensing with a single electron spin using real-time adaptive measurements},
	author={Bonato, Cristian and Blok, Machiel S and Dinani, Hossein T and Berry, Dominic W and Markham, Matthew L and Twitchen, Daniel J and Hanson, Ronald},
	journal={Nature nanotechnology},
	volume={11},
	number={3},
	pages={247--252},
	year={2016},
	publisher={Nature Publishing Group UK London}
}

@article{wang2019heisenberg,
	title={Heisenberg-limited single-mode quantum metrology in a superconducting circuit},
	author={Wang, Weiting and Wu, Yukai and Ma, Yuwei and Cai, Weizhou and Hu, Ling and Mu, Xianghao and Xu, Yuan and Chen, Zi-Jie and Wang, Haiyan and Song, YP and others},
	journal={Nature communications},
	volume={10},
	number={1},
	pages={4382},
	year={2019},
	publisher={Nature Publishing Group UK London}
}

@article{mihailescu2024quantum,
	title={Quantum sensing with nanoelectronics: Fisher information for an adiabatic perturbation},
	author={Mihailescu, George and Kiely, Anthony and Mitchell, Andrew K},
	journal={arXiv e-prints},
	pages={arXiv--2406},
	year={2024}
}

@article{petropoulos2024nanoscale,
	title={Nanoscale single-electron box with a floating lead for quantum sensing: Modeling and device characterization},
	author={Petropoulos, Nikolaos and Wu, Xutong and Sokolov, Andrii and Giounanlis, Panagiotis and Bashir, Imran and Mitchell, Andrew K and Asker, Mike and Leipold, Dirk and Staszewski, Robert B and Blokhina, Elena},
	journal={Applied Physics Letters},
	volume={124},
	number={17},
	year={2024},
	publisher={AIP Publishing}
}

@article{zhou2025entanglement,
	title={Entanglement-enhanced nanoscale single-spin sensing},
	author={Zhou, Xu and Wang, Mengqi and Ye, Xiangyu and Sun, Haoyu and Guo, Yuhang and Han, Shuo and Chai, Zihua and Ji, Wentao and Xia, Kangwei and Shi, Fazhan and others},
	journal={Nature},
	volume={647},
	number={8091},
	pages={883--888},
	year={2025},
	publisher={Nature Publishing Group UK London}
}

@article{hui2026ultrathin,
	title={Ultrathin Fluorescent Nanodiamond Films for Nanoscale Quantum Sensing in Operando Semiconductor Devices},
	author={Hui, Yuen Yung and Tsui, Yi-Mu and Tang, Yi-Xiu and Chang, Huan-Cheng},
	journal={Advanced Functional Materials},
	volume={36},
	number={10},
	pages={e13406},
	year={2026},
	publisher={Wiley Online Library}
}

@article{gaikwad2024entanglement,
	title={Entanglement assisted probe of the non-Markovian to Markovian transition in open quantum system dynamics},
	author={Gaikwad, Chandrashekhar and Kowsari, Daria and Brame, Carson and Song, Xingrui and Zhang, Haimeng and Esposito, Martina and Ranadive, Arpit and Cappelli, Giulio and Roch, Nicolas and Levenson-Falk, Eli M and others},
	journal={Physical Review Letters},
	volume={132},
	number={20},
	pages={200401},
	year={2024},
	publisher={APS}
}

@article{li2026quantum,
	title={Quantum Mpemba Effect in Non-Equilibrium Quantum Thermometry},
	author={Li, Zi-Shen and Yang, Yuxiang},
	journal={arXiv preprint arXiv:2604.14740},
	year={2026}
}

@article{correa2015individual,
	title={Individual quantum probes for optimal thermometry},
	author={Correa, Luis A and Mehboudi, Mohammad and Adesso, Gerardo and Sanpera, Anna},
	journal={Physical review letters},
	volume={114},
	number={22},
	pages={220405},
	year={2015},
	publisher={APS}
}

@article{cavina2018bridging,
	title={Bridging thermodynamics and metrology in nonequilibrium quantum thermometry},
	author={Cavina, Vasco and Mancino, Luca and De Pasquale, Antonella and Gianani, Ilaria and Sbroscia, Marco and Booth, Robert I and Roccia, Emanuele and Raimondi, Roberto and Giovannetti, Vittorio and Barbieri, Marco},
	journal={Physical Review A},
	volume={98},
	number={5},
	pages={050101},
	year={2018},
	publisher={APS}
}

@article{aiache2024non,
	title={Non-Markovian enhancement of nonequilibrium quantum thermometry},
	author={Aiache, Y and Seida, C and El Anouz, K and El Allati, Abderrahim},
	journal={Physical Review E},
	volume={110},
	number={2},
	pages={024132},
	year={2024},
	publisher={APS}
}

@article{alipour2014quantum,
	title={Quantum metrology in open systems: dissipative Cram{\'e}r-Rao bound},
	author={Alipour, S and Mehboudi, M and Rezakhani, AT},
	journal={Physical review letters},
	volume={112},
	number={12},
	pages={120405},
	year={2014},
	publisher={APS}
}

@article{montenegro2025quantum,
	title={Quantum metrology and sensing with many-body systems},
	author={Montenegro, Victor and Mukhopadhyay, Chiranjib and Yousefjani, Rozhin and Sarkar, Saubhik and Mishra, Utkarsh and Paris, Matteo GA and Bayat, Abolfazl},
	journal={Physics Reports},
	volume={1134},
	pages={1--62},
	year={2025},
	publisher={Elsevier}
}

@article{saleem2023optimal,
	title={Optimal time for sensing in open quantum systems},
	author={Saleem, Zain H and Shaji, Anil and Gray, Stephen K},
	journal={Physical Review A},
	volume={108},
	number={2},
	pages={022413},
	year={2023},
	publisher={APS}
}

@article{gorecki2025interplay,
	title={Interplay between time and energy in bosonic noisy quantum metrology},
	author={G{\'o}recki, Wojciech and Albarelli, Francesco and Felicetti, Simone and Di Candia, Roberto and Maccone, Lorenzo},
	journal={PRX Quantum},
	volume={6},
	number={2},
	pages={020351},
	year={2025},
	publisher={APS}
}

@article{hotter2024combining,
	title={Combining critical and quantum metrology},
	author={Hotter, Christoph and Ritsch, Helmut and Gietka, Karol},
	journal={Physical Review Letters},
	volume={132},
	number={6},
	pages={060801},
	year={2024},
	publisher={APS}
}

@article{xiao2026observation,
	title={Observation of Criticality-Enhanced Quantum Sensing in Nonunitary Quantum Walks},
	author={Xiao, Lei and Sarkar, Saubhik and Wang, Kunkun and Bayat, Abolfazl and Xue, Peng},
	journal={Physical Review Letters},
	volume={136},
	number={6},
	pages={060802},
	year={2026},
	publisher={APS}
}

@article{mann2025quantum,
	title={Quantum error-corrected non-Markovian metrology},
	author={Mann, Zachary and Cao, Ningping and Laflamme, Raymond and Zhou, Sisi},
	journal={PRX Quantum},
	volume={6},
	number={3},
	pages={030321},
	year={2025},
	publisher={APS}
}

@article{huang2024entanglement,
	title={Entanglement-enhanced quantum metrology: From standard quantum limit to Heisenberg limit},
	author={Huang, Jiahao and Zhuang, Min and Lee, Chaohong},
	journal={Applied Physics Reviews},
	volume={11},
	number={3},
	year={2024},
	publisher={AIP Publishing}
}

@article{solanki2025universal,
	title={Universal relaxation speedup in open quantum systems through transient conditional and unconditional resetting},
	author={Solanki, Parvinder and Lesanovsky, Igor and Perfetto, Gabriele},
	journal={arXiv preprint arXiv:2512.10005},
	year={2025}
}

@article{rose2018spectral,
	title={Spectral properties of simple classical and quantum reset processes},
	author={Rose, Dominic C and Touchette, Hugo and Lesanovsky, Igor and Garrahan, Juan P},
	journal={Physical Review E},
	volume={98},
	number={2},
	pages={022129},
	year={2018},
	publisher={APS}
}

@article{mukherjee2018quantum,
	title={Quantum dynamics with stochastic reset},
	author={Mukherjee, B and Sengupta, K and Majumdar, Satya N},
	journal={Physical Review B},
	volume={98},
	number={10},
	pages={104309},
	year={2018},
	publisher={APS}
}

@article{perfetto2022thermodynamics,
	title={Thermodynamics of quantum-jump trajectories of open quantum systems subject to stochastic resetting},
	author={Perfetto, Gabriele and Carollo, Federico and Lesanovsky, Igor},
	journal={SciPost Physics},
	volume={13},
	number={4},
	pages={079},
	year={2022}
}

@article{kuffer2025sensing,
	title={Sensing Out-of-Equilibrium and Quantum Non-Gaussian environments via induced Time-Reversal Symmetry Breaking on the quantum-probe dynamics},
	author={Kuffer, Martin and Zwick, Analia and {\'A}lvarez, Gonzalo A},
	journal={PRX Quantum},
	volume={6},
	number={2},
	pages={020320},
	year={2025},
	publisher={APS}
}

@article{rosenberg2026witnessing,
	title={Witnessing nonstationary and non-Markovian environments with a quantum sensor},
	author={Rosenberg, John W and Kuffer, Mart{\'\i}n and Zohar, Inbar and St{\"o}hr, Rainer and Denisenko, Andrej and Zwick, Analia and {\'A}lvarez, Gonzalo A and Finkler, Amit},
	journal={Physical Review Applied},
	volume={25},
	number={1},
	pages={014049},
	year={2026},
	publisher={APS}
}

@article{alushi2025collective,
	title={Collective quantum enhancement in critical quantum sensing},
	author={Alushi, Uesli and Coppo, Alessandro and Brosco, Valentina and Di Candia, Roberto and Felicetti, Simone},
	journal={Communications Physics},
	volume={8},
	number={1},
	pages={74},
	year={2025},
	publisher={Nature Publishing Group UK London}
}

@article{alushi2024optimality,
	title={Optimality and noise resilience of critical quantum sensing},
	author={Alushi, Uesli and G{\'o}recki, Wojciech and Felicetti, Simone and Di Candia, Roberto},
	journal={Physical Review Letters},
	volume={133},
	number={4},
	pages={040801},
	year={2024},
	publisher={APS}
}

@article{beaulieu2025criticality,
	title={Criticality-enhanced quantum sensing with a parametric superconducting resonator},
	author={Beaulieu, Guillaume and Minganti, Fabrizio and Frasca, Simone and Scigliuzzo, Marco and Felicetti, Simone and Di Candia, Roberto and Scarlino, Pasquale},
	journal={PRX Quantum},
	volume={6},
	number={2},
	pages={020301},
	year={2025},
	publisher={APS}
}

@article{sarkar2025exponentially,
	title={Exponentially-enhanced quantum sensing with many-body phase transitions},
	author={Sarkar, Saubhik and Bayat, Abolfazl and Bose, Sougato and Ghosh, Roopayan},
	journal={Nature Communications},
	volume={16},
	number={1},
	pages={5159},
	year={2025},
	publisher={Nature Publishing Group UK London}
}

@article{long2022entanglement,
	title={Entanglement-enhanced quantum metrology in colored noise by quantum Zeno effect},
	author={Long, Xinyue and He, Wan-Ting and Zhang, Na-Na and Tang, Kai and Lin, Zidong and Liu, Hongfeng and Nie, Xinfang and Feng, Guanru and Li, Jun and Xin, Tao and others},
	journal={Physical Review Letters},
	volume={129},
	number={7},
	pages={070502},
	year={2022},
	publisher={APS}
}

@article{wang2018entanglement,
	title={Entanglement-enhanced quantum metrology in a noisy environment},
	author={Wang, Kunkun and Wang, Xiaoping and Zhan, Xiang and Bian, Zhihao and Li, Jian and Sanders, Barry C and Xue, Peng},
	journal={Physical Review A},
	volume={97},
	number={4},
	pages={042112},
	year={2018},
	publisher={APS}
}

@article{zou2025enhancing,
	title={Enhancing quantum metrology by quantum resonance dynamics},
	author={Zou, Zhixing and Gong, Jiangbin and Chen, Weitao},
	journal={Physical Review Letters},
	volume={134},
	number={23},
	pages={230802},
	year={2025},
	publisher={APS}
}

@article{zhai2023control,
	title={Control-enhanced quantum metrology under Markovian noise},
	author={Zhai, Yue and Yang, Xiaodong and Tang, Kai and Long, Xinyue and Nie, Xinfang and Xin, Tao and Lu, Dawei and Li, Jun},
	journal={Physical Review A},
	volume={107},
	number={2},
	pages={022602},
	year={2023},
	publisher={APS}
}

@article{rahim2024entanglement,
	title={Entanglement-enhanced optimal quantum metrology},
	author={Rahim, Muhammad Talha and Al-Kuwari, Saif and Ali, Asad},
	journal={arXiv preprint arXiv:2411.04022},
	year={2024}
}

@article{demkowicz2014using,
	title={Using entanglement against noise in quantum metrology},
	author={Demkowicz-Dobrza{\'n}ski, Rafal and Maccone, Lorenzo},
	journal={Physical review letters},
	volume={113},
	number={25},
	pages={250801},
	year={2014},
	publisher={APS}
}

@article{huang2018noise,
	title={Noise-dependent optimal strategies for quantum metrology},
	author={Huang, Zixin and Macchiavello, Chiara and Maccone, Lorenzo},
	journal={Physical Review A},
	volume={97},
	number={3},
	pages={032333},
	year={2018},
	publisher={APS}
}

@article{cao2026cross,
	title={Cross-scale Raman thermometry: Bridging equilibrium-to-non-equilibrium thermal dynamics from micro-devices to single-molecule junctions},
	author={Cao, Yu and Tan, Xiaomin and Liu, Dunmei and Zheng, Zhexiong and Wang, Weicheng and Zeng, Yanqing and Li, Jing and Liu, Junyang and Gao, Mingbin and Hong, Wenjing},
	journal={Nano Research},
	year={2026},
	publisher={清华大学出版社}
}

@article{saha2026precision,
	title={Precision Enhancement in Transient Quantum Thermometry: Cold-Probe Bias and Its Removal},
	author={Saha, Debarupa and Sen, Ujjwal},
	journal={arXiv preprint arXiv:2601.15237},
	year={2026}
}

@article{neumann2013high,
	title={High-precision nanoscale temperature sensing using single defects in diamond},
	author={Neumann, Philipp and Jakobi, Ingmar and Dolde, Florian and Burk, Christian and Reuter, Rolf and Waldherr, Gerald and Honert, Jan and Wolf, Thomas and Brunner, Andreas and Shim, Jeong Hyun and others},
	journal={Nano letters},
	volume={13},
	number={6},
	pages={2738--2742},
	year={2013},
	publisher={ACS Publications}
}

@article{mukhopadhyay2025current,
	title={Current trends in global quantum metrology},
	author={Mukhopadhyay, Chiranjib and Montenegro, Victor and Bayat, Abolfazl},
	journal={Journal of Physics A: Mathematical and Theoretical},
	volume={58},
	number={6},
	pages={063001},
	year={2025},
	publisher={IOP Publishing}
}

@article{gebbia2020two,
	title={Two-qubit quantum probes for the temperature of an Ohmic environment},
	author={Gebbia, Francesca and Benedetti, Claudia and Benatti, Fabio and Floreanini, Roberto and Bina, Matteo and Paris, Matteo GA},
	journal={Physical Review A},
	volume={101},
	number={3},
	pages={032112},
	year={2020},
	publisher={APS}
}

@article{glatthard2022bending,
	title={Bending the rules of low-temperature thermometry with periodic driving},
	author={Glatthard, Jonas and Correa, Luis A},
	journal={Quantum},
	volume={6},
	pages={705},
	year={2022},
	publisher={Verein zur F{\"o}rderung des Open Access Publizierens in den Quantenwissenschaften}
}

@article{plodzien2018few,
	title={Few-fermion thermometry},
	author={P{\l}odzie{\'n}, Marcin and Demkowicz-Dobrza{\'n}ski, Rafa{\l} and Sowi{\'n}ski, Tomasz},
	journal={Physical Review A},
	volume={97},
	number={6},
	pages={063619},
	year={2018},
	publisher={APS}
}

@article{sekatski2022optimal,
	title={Optimal nonequilibrium thermometry in Markovian environments},
	author={Sekatski, Pavel and Perarnau-Llobet, Mart{\'\i}},
	journal={Quantum},
	volume={6},
	pages={869},
	year={2022},
	publisher={Verein zur F{\"o}rderung des Open Access Publizierens in den Quantenwissenschaften}
}

@article{xu2023non,
	title={Non-Markovian enhanced temperature sensing in a dipolar Bose-Einstein condensate},
	author={Xu, Lan and Yuan, Ji-Bing and Tang, Shi-Qing and Wu, Wei and Tan, Qing-Shou and Kuang, Le-Man},
	journal={Physical Review A},
	volume={108},
	number={2},
	pages={022608},
	year={2023},
	publisher={APS}
}

@article{albarelli2023invasiveness,
	title={Invasiveness of nonequilibrium pure-dephasing quantum thermometry},
	author={Albarelli, Francesco and Paris, Matteo GA and Vacchini, Bassano and Smirne, Andrea},
	journal={Physical Review A},
	volume={108},
	number={6},
	pages={062421},
	year={2023},
	publisher={APS}
}

@article{frazao2024coherence,
	title={Coherence-enhanced single-qubit thermometry out of equilibrium},
	author={Fraz{\~a}o, Gon{\c{c}}alo and Pezzutto, Marco and Omar, Yasser and Zambrini Cruzeiro, Emmanuel and Gherardini, Stefano},
	journal={Entropy},
	volume={26},
	number={7},
	pages={568},
	year={2024},
	publisher={MDPI}
}

@article{aiache2025nonequilibrium,
	title={Nonequilibrium quantum thermometry with noncommutative system-bath couplings},
	author={Aiache, Youssef and Allati, Abderrahim El and Demir, {\.I}lkay and Anouz, Khadija El},
	journal={Physical Review A},
	volume={112},
	number={6},
	pages={062229},
	year={2025},
	publisher={APS}
}

@article{brenes2023multispin,
	title={Multispin probes for thermometry in the strong-coupling regime},
	author={Brenes, Marlon and Segal, Dvira},
	journal={Physical Review A},
	volume={108},
	number={3},
	pages={032220},
	year={2023},
	publisher={APS}
}

@article{ravell2024strongly,
	title={Strongly coupled fermionic probe for nonequilibrium thermometry},
	author={Ravell Rodr{\'\i}guez, Ricard and Mehboudi, Mohammad and Horodecki, Micha{\l} and Perarnau-Llobet, Marti},
	journal={New journal of physics},
	volume={26},
	number={1},
	pages={013046},
	year={2024},
	publisher={IOP Publishing}
}

@article{xie2026direct,
	title={Direct temperature readout in nonequilibrium quantum thermometry},
	author={Xie, Yan and Liu, Junjie},
	journal={arXiv preprint arXiv:2601.07198},
	year={2026}
}

@article{campaioli2024quantum,
	title={Quantum master equations: Tips and tricks for quantum optics, quantum computing, and beyond},
	author={Campaioli, Francesco and Cole, Jared H and Hapuarachchi, Harini},
	journal={PRX Quantum},
	volume={5},
	number={2},
	pages={020202},
	year={2024},
	publisher={APS}
}

@article{boixo2007generalized,
	title={Generalized limits for single-parameter quantum estimation},
	author={Boixo, Sergio and Flammia, Steven T and Caves, Carlton M and Geremia, John M},
	journal={Physical review letters},
	volume={98},
	number={9},
	pages={090401},
	year={2007},
	publisher={APS}
}

@article{roy2008exponentially,
	title={Exponentially enhanced quantum metrology},
	author={Roy, SM and Braunstein, Samuel L},
	journal={Physical review letters},
	volume={100},
	number={22},
	pages={220501},
	year={2008},
	publisher={APS}
}

@article{lindblad1976generators,
	title={On the generators of quantum dynamical semigroups},
	author={Lindblad, Goran},
	journal={Communications in mathematical physics},
	volume={48},
	number={2},
	pages={119--130},
	year={1976},
	publisher={Springer}
}

@article{zhong2013fisher,
	title={Fisher information under decoherence in Bloch representation},
	author={Zhong, Wei and Sun, Zhe and Ma, Jian and Wang, Xiaoguang and Nori, Franco},
	journal={Physical Review A—Atomic, Molecular, and Optical Physics},
	volume={87},
	number={2},
	pages={022337},
	year={2013},
	publisher={APS}
}

@article{buvzek1999optimal,
	title={Optimal quantum clocks},
	author={Bu{\v{z}}ek, Vladimir and Derka, Radoslav and Massar, Serge},
	journal={Physical review letters},
	volume={82},
	number={10},
	pages={2207},
	year={1999},
	publisher={APS}
}

@article{faist2023time,
	title={Time-energy uncertainty relation for noisy quantum metrology},
	author={Faist, Philippe and Woods, Mischa P and Albert, Victor V and Renes, Joseph M and Eisert, Jens and Preskill, John},
	journal={PRX Quantum},
	volume={4},
	number={4},
	pages={040336},
	year={2023},
	publisher={APS}
}

@article{prech2025optimal,
	title={Optimal time estimation and the clock uncertainty relation for stochastic processes},
	author={Prech, Kacper and Landi, Gabriel T and Meier, Florian and Nurgalieva, Nuriya and Potts, Patrick P and Silva, Ralph and Mitchison, Mark T},
	journal={Physical Review X},
	volume={15},
	number={3},
	pages={031068},
	year={2025},
	publisher={APS}
}

@article{vsafranek2017discontinuities,
	title={Discontinuities of the quantum Fisher information and the Bures metric},
	author={{\v{S}}afr{\'a}nek, Dominik},
	journal={Physical Review A},
	volume={95},
	number={5},
	pages={052320},
	year={2017},
	publisher={APS}
}

\end{document}

\title{Supplemental Material for ``Reset-Enhanced Dissipative Quantum Metrology''}
	
	\author{Silan Li}
	\affiliation{School of Physics and Electronics, Hunan University, Changsha 410082, China}
	
	\author{Qingjun Tong}
	\email{tongqj@hnu.edu.cn}
	\affiliation{School of Physics and Electronics, Hunan University, Changsha 410082, China}
	
	\author{Ruicheng Bao}
	\email{ruicheng@g.ecc.u-tokyo.ac.jp}
	\affiliation{Department of Physics, Graduate School of Science,The University of Tokyo, Hongo, Bunkyo-ku, Tokyo 113-0033, Japan}
	
	\maketitle

	\section{General Fisher-information response identity}
	For reference we collect the notation used below. 
	For a smooth reset-dependent state  \(\varrho_{\theta,r}\), the symmetric logarithmic derivative (SLD) \(\mathbb{L}_{\theta,r}\) is defined by
	\begin{equation}
		\partial_\theta\varrho_{\theta,r}=\frac12\{\varrho_{\theta,r},\mathbb{L}_{\theta,r}\}.
		\label{eq:S_sld_def}
	\end{equation}
	The Fisher information is
	\begin{equation}
		F[\varrho_{\theta,r}]=\Tr[(\partial_\theta\varrho_{\theta,r})\mathbb{L}_{\theta,r}]
		=\Tr[\varrho_{\theta,r}\mathbb{L}_{\theta,r}^2].
		\label{eq:S_F_def}
	\end{equation}
	As derived in the End Matter of the main text, its exact reset response is
	\begin{equation}
		\partial_rF[\varrho_{\theta,r}]=2\Tr[(\partial_\theta\partial_r\varrho_{\theta,r})\mathbb{L}_{\theta,r}]-\Tr[(\partial_r\varrho_{\theta,r})\mathbb{L}_{\theta,r}^2].
		\label{eq:S_response}
	\end{equation}
	This identity applies to the steady state
	\(\varrho_{\theta,r}=\rho_{\theta,r}^{\rm ss}\)
	and to the transient state
	\(\varrho_{\theta,r}=\rho_{\theta,r}(t)\) at fixed readout time \(t\).
	For rank-deficient states, \(\mathbb{L}_{\theta,r}\) is understood through the Moore--Penrose inverse of the corresponding Jordan map on the locally smooth support of \(\varrho_{\theta,r}\), provided \(F[\varrho_{\theta,r}]\) is differentiable in \(r\).
	
	A strict local small-reset enhancement condition is therefore
	\begin{equation}
		\left.
		\partial_rF[\varrho_{\theta,r}]
		\right|_{r=0}
		>0.
		\label{eq:S_local_condition}
	\end{equation}
	The remaining sections verify this condition explicitly for the plotted parameter regimes.

	\section{Reset steady state and transient modal solution}
	
	Assume that the unreset Lindbladian is diagonalizable and has a unique steady state:
	\begin{equation}
		\Lind_\theta(R_j)=\lambda_jR_j,\quad\Lind_\theta^\dagger( L_j)=\lambda_j^* L_j,\quad\Tr( L_j^\dagger R_k)=\delta_{jk},
		\label{eq:S_modes}
	\end{equation}
	with $\lambda_1=0$, $R_1=\rho_\theta^{\rm ss}$, and $0\ge\operatorname{Re}\lambda_2\ge\operatorname{Re}\lambda_3\ge\cdots$. For the reset target,
	\begin{equation}
		\rho_\delta=\rho_\theta^{\rm ss}+\sum_{j\ge2}d_j(\theta)R_j(\theta),\quad d_j(\theta)=\Tr[ L_j^\dagger(\theta)\rho_\delta].
		\label{eq:S_dj}
	\end{equation}
	The reset master equation for normalized states is
	\begin{equation}
		\dot\rho=\Lind_\theta(\rho)+r(\rho_\delta-\rho).
		\label{eq:S_resetME}
	\end{equation}
	Seeking the reset steady state as $\rho_{\theta,r}^{\rm ss}=\rho_\theta^{\rm ss}+\sum_{j\ge2}a_jR_j$, Eq.~\eqref{eq:S_resetME} gives
	\begin{equation}
		0=\sum_{j\ge2}(\lambda_ja_j+rd_j-ra_j)R_j,\quad a_j=-\frac{rd_j}{\lambda_j-r}.
	\end{equation}
	Hence
	\begin{equation}
		\rho_{\theta,r}^{\rm ss}=\rho_\theta^{\rm ss}-\sum_{j\ge2}\frac{r\,d_j(\theta)}{\lambda_j(\theta)-r}R_j(\theta).
		\label{eq:S_ssmodal}
	\end{equation}
	For small $r$,
	\begin{equation}
		\rho_{\theta,r}^{\rm ss}=\rho_\theta^{\rm ss}+r\rho_{1,\theta}^{\rm ss}+O(r^2),\quad\rho_{1,\theta}^{\rm ss}=-\sum_{j\ge2}\frac{d_j(\theta)}{\lambda_j(\theta)}R_j(\theta).
		\label{eq:S_rho1ss}
	\end{equation}
	Using Eq.~\eqref{eq:S_response}, the strict first-order sufficient condition for local steady-state enhancement is
	\begin{equation}
		\left.\partial_rF[\rho_{\theta,r}^{\rm ss}]\right|_{r=0}=2\Tr[(\partial_\theta\rho_{1,\theta}^{\rm ss})\mathbb{L}_\theta^{(0)}]-\Tr[\rho_{1,\theta}^{\rm ss}(\mathbb{L}_\theta^{(0)})^2]>0.
		\label{eq:S_sscriterion}
	\end{equation}

	For a temporary reset pulse, write the initial state as
	\begin{equation}
		\rho_0=\rho_\theta^{\rm ss}+\sum_{k\ge2}c_kR_k,\quad c_k=\Tr[ L_k^\dagger\rho_0].
	\end{equation}
	During the pulse $0\le t\le t_s$, the exact modal solution is
	\begin{align}
		\rho_{\theta,r}(t)&=\rho_\theta^{\rm ss}+\sum_{k\ge2}c_k^{\,r}(t)e^{\lambda_kt}R_k,
		\label{eq:S_transientmodal_a}\\
		c_k^{\,r}(t)&=\left(c_k-\frac{rd_k}{r-\lambda_k}\right)e^{-rt}+\frac{rd_k}{r-\lambda_k}e^{-\lambda_kt}.
		\label{eq:S_transientmodal_b}
	\end{align}
	After switch-off, $c_k^{\,r}(t_s)$ is fixed and the subsequent evolution proceeds under the original spectrum.
	
	At a fixed readout time, Eq.~\eqref{eq:S_response} gives
	\begin{equation}
		\left.\partial_rF[\rho_r(t;\theta)]\right|_{r=0}=2\Tr[(\partial_\theta\left.\partial_r\rho_{\theta,r}(t)\right|_{r=0})\mathbb{L}_t^{(0)}]
		-\Tr[\left.\partial_r\rho_{\theta,r}(t)\right|_{r=0}(\mathbb{L}_t^{(0)})^2]>0,
		\label{eq:smallrcriterion2}
	\end{equation}
	where \(\mathbb{L}_t^{(0)}\) is the SLD of the unreset transient state \(\rho_{\theta,0}(t)\). 
	A positive value is a strict first-order sufficient condition for local transient enhancement.
	
	\section{Exact steady-state thermometry in a reset Davies qubit}
	We consider a qubit with Hamiltonian $H=(\omega/2)\sigma_z$ and thermal Davies generator
	\begin{equation}
		\mathcal{L}_\beta(\rho)
		=-i[H,\rho]+
		\gamma_\downarrow\left(\sigma_-\rho\sigma_+-\frac{1}{2}\{\sigma_+\sigma_-,\rho\}\right)
		+
		\gamma_\uparrow\left(\sigma_+\rho\sigma_--\frac{1}{2}\{\sigma_-\sigma_+,\rho\}\right),
		\label{eq:S_davies_generator}
	\end{equation}
	with bosonic rates
	\begin{equation}
		\gamma_\uparrow(\beta)=\gamma n(\beta),\quad
		\gamma_\downarrow(\beta)=\gamma[n(\beta)+1],\quad
		n(\beta)=\frac{1}{e^{\beta\omega}-1}.
		\label{eq:S_bosonic_rates}
	\end{equation}
	
	The estimated parameter is $\beta$, and the reset target is the ground state $|0\rangle=(0,1)^{\rm T}$, so $p_\delta=0$. In the energy basis, the steady state is diagonal. Let $p_{\beta,r}^{\rm ss}$ be the excited-state population. Then
	\begin{equation}
		p_{\beta,r}^{\rm ss}=\frac{\gamma}{\gamma(e^{\beta\omega}+1)+r(e^{\beta\omega}-1)}.
		\label{eq:S_p_ss}
	\end{equation}
	Since the eigenbasis is $\beta$-independent, the Fisher information reduces to
	\begin{equation}
		F^{\rm ss}(\beta;r)=\frac{(\partial_\beta p_{\beta,r}^{\rm ss})^2}{p_{\beta,r}^{\rm ss}(1-p_{\beta,r}^{\rm ss})}.
		\label{eq:S_Fss_binary}
	\end{equation}
	Define
	\begin{equation}
		q=e^{\beta\omega},\quad s=\frac{r}{\gamma},\quad B=(1+s)q-s.
	\end{equation}
	Equation~\eqref{eq:S_p_ss} becomes
	\begin{equation}
		p_{\beta,r}^{\rm ss}=\frac{1}{(1+s)q-s+1}=\frac{1}{B+1}.
		\label{eq:S_p_B}
	\end{equation}
	Therefore
	\begin{equation}
		1-p_{\beta,r}^{\rm ss}=\frac{B}{B+1},\quad p_{\beta,r}^{\rm ss}(1-p_{\beta,r}^{\rm ss})=\frac{B}{(B+1)^2}.
		\label{eq:S_p_variance}
	\end{equation}
	Moreover, because $\partial_\beta q=\omega q$,
	\begin{equation}
		\partial_\beta p_{\beta,r}^{\rm ss}=-\frac{\partial_\beta B}{(B+1)^2}=-\frac{(1+s)\omega q}{(B+1)^2}.
		\label{eq:S_dp_dbeta}
	\end{equation}
	Substituting Eqs.~\eqref{eq:S_p_variance} and \eqref{eq:S_dp_dbeta} into Eq.~\eqref{eq:S_Fss_binary} gives
	\begin{align}
		F^{\rm ss}(\beta;s) =\frac{\omega^2 q^2(1+s)^2/(B+1)^4}{B/(B+1)^2}
		=\omega^2\frac{q^2(1+s)^2}{B(B+1)^2}.
	\end{align}
	Restoring $B=(1+s)q-s$, we obtain
	\begin{equation}
		F^{\rm ss}(\beta;s)=\omega^2\frac{q^2(1+s)^2}{[(1+s)q-s]\bigl[(1+s)q-s+1\bigr]^2}.
		\label{eq:S_Fss}
	\end{equation}
	The unreset value follows by setting $s=0$:
	\begin{equation}
		F^{\rm ss}(\beta;0)=\omega^2\frac{q}{(1+q)^2}.
		\label{eq:S_Fss0}
	\end{equation}
	
	The exact gain region can be obtained analytically from the inequality $F^{\rm ss}(\beta;s)>F^{\rm ss}(\beta;0)$. Using Eqs.~\eqref{eq:S_Fss} and \eqref{eq:S_Fss0}, this inequality is equivalent to
	\begin{equation}
		\frac{q^2(1+s)^2}{[(1+s)q-s][(1+s)q-s+1]^2}>\frac{q}{(q+1)^2}.
		\label{eq:S_gain_ineq_raw}
	\end{equation}
	For $q>1$ and $s\ge0$, all denominators are positive. After clearing denominators, Eq.~\eqref{eq:S_gain_ineq_raw} becomes
	\begin{equation}
		q(1+s)^2(q+1)^2>[(1+s)q-s][(1+s)q-s+1]^2.
		\label{eq:S_gain_ineq_clear}
	\end{equation}
	Moving all terms to the left gives
	\begin{equation}
		q(1+s)^2(q+1)^2-[(1+s)q-s][(1+s)q-s+1]^2=-s\,Q(q,s),
		\label{eq:S_gain_Q_def}
	\end{equation}
	where
	\begin{equation}
		Q(q,s)=(q-1)^3s^2+2(q^3-3q^2-q+1)s+(q^3-3q^2-5q-1).
		\label{eq:S_Q_poly}
	\end{equation}
	Therefore, for physical reset $s>0$,
	\begin{equation}
		F^{\rm ss}(\beta;s)>F^{\rm ss}(\beta;0)\iff Q(q,s)<0.
		\label{eq:S_Q_condition}
	\end{equation}
	
	Since $q>1$, the coefficient of $s^2$ in $Q(q,s)$ is positive. The discriminant is
	\begin{equation}
		\Delta=4(q^3-3q^2-q+1)^2-4(q-1)^3(q^3-3q^2-5q-1)=16q(q^2+q-1)>0.
		\label{eq:S_Q_discriminant}
	\end{equation}
	Hence $Q(q,s)=0$ has two real roots,
	\begin{subequations}
		\begin{align}
			s_-(q)&=\frac{-q^3+3q^2+q-1-2\sqrt{q(q^2+q-1)}}{(q-1)^3},\\
			s_+(q)&=\frac{-q^3+3q^2+q-1+2\sqrt{q(q^2+q-1)}}{(q-1)^3}.
		\end{align}
		\label{eq:S_sroots}
	\end{subequations}
	
	\begin{figure}[t]
		\centering
		\includegraphics[width=0.88\linewidth]{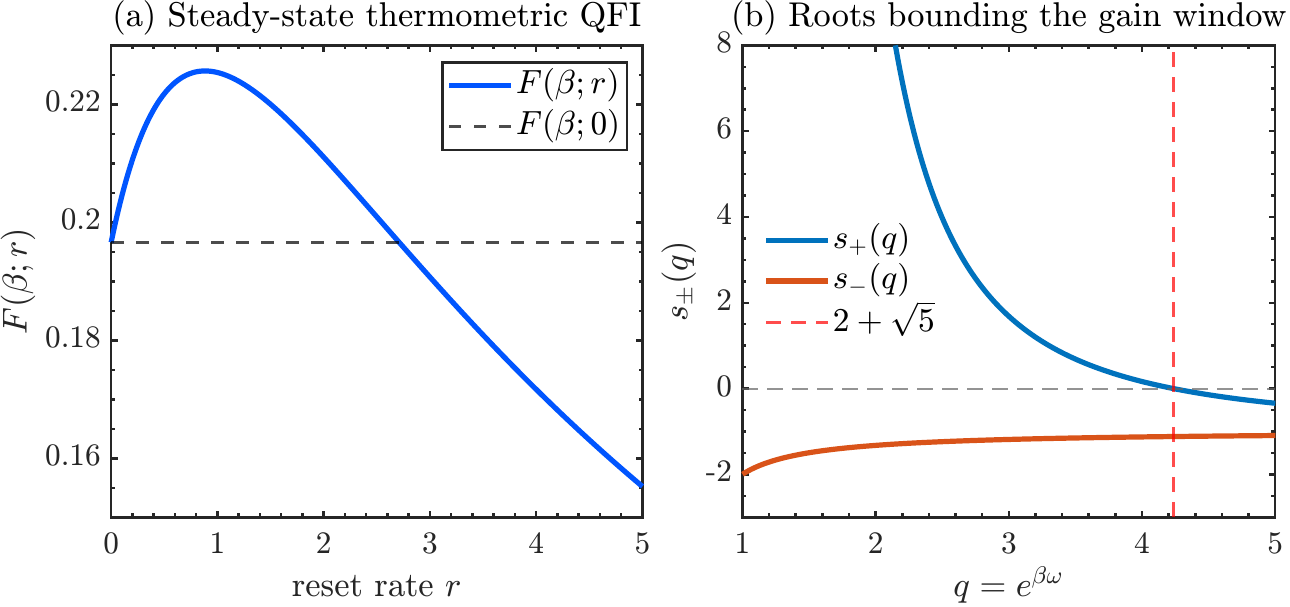}
		\caption{
			Steady-state thermometric gain window for the reset Davies qubit.
			(a) Fisher information \(F^{\rm ss}(\beta;r)\) at \(\beta=\omega=\gamma=1\), with the dashed line marking \(F^{\rm ss}(\beta;0)\).
			The maximum occurs at the analytic optimum \(r_{\rm opt}\).
			(b) Roots \(s_\pm(q)\) of \(Q(q,s)=0\). The lower branch is negative for \(q>1\), while the upper branch is positive only for \(1<q<2+\sqrt5\), giving the physical gain window \(0<s<s_+(q)\).
		}
		\label{fig:S1}
	\end{figure}
	
	For all \(q>1\), the lower root satisfies \(s_-(q)<0\), while \(s_+(q)>0\) only for \(1<q<2+\sqrt5\), as illustrated in Fig.~\ref{fig:S1}. Thus the physical gain region is controlled only by \(s_+(q)\). Moreover,
	\begin{equation}
		s_+(q)>0\iff q^3-3q^2-5q-1<0\iff (q+1)(q^2-4q-1)<0.
	\end{equation}
	Since $q>1$, this gives $1<q<2+\sqrt5$. Consequently,
	\begin{equation}
		F^{\rm ss}(\beta;s)>F^{\rm ss}(\beta;0)\iff 0<\beta\omega<\ln(2+\sqrt5),\quad 0<s<s_+(e^{\beta\omega})
		\label{eq:S_exactwindow}
	\end{equation}
	
	We next determine the optimal reset rate. Write Eq.~\eqref{eq:S_Fss} as
	\begin{equation}
		F^{\rm ss}(\beta;s)=\omega^2q^2\frac{(1+s)^2}{A(s)[A(s)+1]^2},\quad A(s)=q+(q-1)s.
		\label{eq:S_F_A}
	\end{equation}
	The constant prefactor $\omega^2q^2$ is irrelevant for maximization. Taking a logarithmic derivative gives
	\begin{equation}
		\partial_s\ln F^{\rm ss}=\frac{2}{1+s}-\frac{A'(s)}{A(s)}-\frac{2A'(s)}{A(s)+1},\quad A'(s)=q-1.
		\label{eq:S_log_derivative}
	\end{equation}
	The steady points satisfy
	\begin{equation}
		\frac{2}{1+s}-\frac{q-1}{A(s)}-\frac{2(q-1)}{A(s)+1}=0.
		\label{eq:S_stationary_eq}
	\end{equation}
	Substituting $A(s)=q+(q-1)s$ and simplifying yields
	\begin{equation}
		(q-1)^2s^2+2(q-1)(q-2)s+(q^2-4q-1)=0.
		\label{eq:S_opt_quad}
	\end{equation}
	The two roots are
	\begin{equation}
		s_1(q)=\frac{2+\sqrt5-q}{q-1},\quad s_2(q)=\frac{2-\sqrt5-q}{q-1}.
		\label{eq:S_opt_roots}
	\end{equation}
	For $q>1$, $s_2(q)<0$ and is unphysical. The remaining steady point is physical only when $s_1(q)>0$, namely when $1<q<2+\sqrt5$. If $q\ge2+\sqrt5$, the maximum over $s\ge0$ is attained at the boundary $s=0$. Thus
	\begin{equation}
		s_{\rm opt}(q)=\max\left[0,\frac{2+\sqrt5-q}{q-1}\right],\quad r_{\rm opt}(\beta)=\gamma\max\left[0,\frac{2+\sqrt5-e^{\beta\omega}}{e^{\beta\omega}-1}\right]
		\label{eq:S_ropt}
	\end{equation}
	
	At the optimum, for $1<q<2+\sqrt5$,
	\begin{equation}
		s_{\rm opt}(q)=\frac{2+\sqrt5-q}{q-1},\quad A(s_{\rm opt})=2+\sqrt5,\quad 1+s_{\rm opt}=\frac{1+\sqrt5}{q-1}.
	\end{equation}
	Substituting these expressions into Eq.~\eqref{eq:S_F_A} gives
	\begin{equation}
		F_{\max}^{\rm ss}(\beta)=\omega^2\frac{5\sqrt5-11}{2}\frac{e^{2\beta\omega}}{(e^{\beta\omega}-1)^2},\quad 1<e^{\beta\omega}<2+\sqrt5.
		\label{eq:S_Fmax_inside}
	\end{equation}
	For $e^{\beta\omega}\ge2+\sqrt5$, the optimum occurs at the boundary $s=0$, so $F_{\max}^{\rm ss}(\beta)=F^{\rm ss}(\beta;0)$. Equivalently,
	\begin{equation}
		F_{\max}^{\rm ss}(\beta)=
		\begin{cases}
			\displaystyle\omega^2\frac{5\sqrt5-11}{2}\frac{e^{2\beta\omega}}{(e^{\beta\omega}-1)^2},&1<e^{\beta\omega}<2+\sqrt5,\\[1.2ex]
			\displaystyle\omega^2\frac{e^{\beta\omega}}{(1+e^{\beta\omega})^2},&e^{\beta\omega}\ge2+\sqrt5.
		\end{cases}
		\label{eq:S_Fmax_piecewise}
	\end{equation}
	See Fig.~\ref{fig:S1} for details.

	\section{Verification of the steady small-reset condition}
	
	We now verify Eq.~\eqref{eq:S_sscriterion} for the Davies qubit. We use the same convention as in the main text,
	\[
	\rho_{\beta,r}^{\rm ss}=
	\begin{pmatrix}
		p_{\beta,r}^{\rm ss}&0\\
		0&1-p_{\beta,r}^{\rm ss}
	\end{pmatrix},
	\qquad
	p_\beta\equiv p_{\beta,0}^{\rm ss}=\frac{1}{q+1},
	\qquad
	q=e^{\beta\omega}.
	\]
	The population relaxation rate at \(r=0\) is
	\[
	\Gamma_\beta=\gamma_\uparrow+\gamma_\downarrow
	=\gamma\frac{q+1}{q-1}.
	\]
	The population right mode and its eigenvalue are
	\[
	R_2=\begin{pmatrix}1&0\\0&-1\end{pmatrix},
	\qquad
	\lambda_2=-\Gamma_\beta .
	\]
	A biorthogonal left mode satisfying
	\(\Tr( L_2^\dagger R_2)=1\) and
	\(\Tr( L_2^\dagger\rho_{\beta,0}^{\rm ss})=0\) is
	\[ \mathbb{L}_2=
	\begin{pmatrix}
		1-p_\beta&0\\
		0&-p_\beta
	\end{pmatrix}.
	\]
	For ground-state reset,
	\[
	\rho_\delta=|0\rangle\langle0|=
	\begin{pmatrix}0&0\\0&1\end{pmatrix},
	\qquad
	d_2=\Tr( L_2^\dagger\rho_\delta)=-p_\beta .
	\]
	The general small-reset expansion
	\[
	\rho_{\beta,r}^{\rm ss}
	=\rho_{\beta,0}^{\rm ss}
	+r\rho_{1,\beta}^{\rm ss}+O(r^2),
	\qquad
	\rho_{1,\beta}^{\rm ss}
	=-\sum_{j\ge2}\frac{d_j}{\lambda_j}R_j
	\]
	therefore gives
	\[
	\rho_{1,\beta}^{\rm ss}
	=-\frac{d_2}{\lambda_2}R_2
	=-\frac{q-1}{\gamma(q+1)^2}R_2
	\equiv a_\beta R_2 .
	\]
	Equivalently,
	\[
	p_{\beta,r}^{\rm ss}=p_\beta+r a_\beta+O(r^2),
	\qquad
	a_\beta=-\frac{q-1}{\gamma(q+1)^2}.
	\]
	This explicitly connects the modal correction \(\rho_{1,\beta}^{\rm ss}\) with the population expansion used in the main text.
	
	The unreset SLD is diagonal:
	\[
	\mathbb{L}_\beta^{(0)}
	=
	\begin{pmatrix}
		-\omega(1-p_\beta)&0\\
		0&\omega p_\beta
	\end{pmatrix}.
	\]
	Using
	\[
	\Tr(R_2\mathbb{L}_\beta^{(0)})=-\omega,\qquad
	\Tr[R_2(\mathbb{L}_\beta^{(0)})^2]=\omega^2\frac{q-1}{q+1},
	\qquad
	\partial_\beta a_\beta=\frac{\omega q(q-3)}{\gamma(q+1)^3},
	\]
	the right-hand side of Eq.~\eqref{eq:S_sscriterion} becomes
	\begin{align}
		2\Tr[(\partial_\beta\rho_{1,\beta}^{\rm ss})\mathbb{L}_\beta^{(0)}]
		-\Tr[\rho_{1,\beta}^{\rm ss}(\mathbb{L}_\beta^{(0)})^2]
		=2(\partial_\beta a_\beta)\Tr(R_2\mathbb{L}_\beta^{(0)})-a_\beta\Tr[R_2(\mathbb{L}_\beta^{(0)})^2]
		=\frac{\omega^2(-q^2+4q+1)}{\gamma(q+1)^3}.
		\label{eq:S_RHS_ss}
	\end{align}
	Direct differentiation of Eq.~\eqref{eq:S_Fss} gives the same result:
	\[
	\left.\partial_s\ln F^{\rm ss}(\beta;s)\right|_{s=0}
	=
	2-\frac{q-1}{q}-2\frac{q-1}{q+1}
	=
	\frac{-q^2+4q+1}{q(q+1)},
	\]
	and since \(s=r/\gamma\),
	\begin{equation}
		\left.\partial_rF^{\rm ss}(\beta;r)\right|_{r=0}
		=
		\frac{\omega^2(-q^2+4q+1)}{\gamma(q+1)^3}.
		\label{eq:S_LHS_ss}
	\end{equation}
	Thus Eqs.~\eqref{eq:S_RHS_ss} and \eqref{eq:S_LHS_ss} are identical, verifying Eq.~\eqref{eq:S_sscriterion} in this model.
	
	The local enhancement condition is $-q^2+4q+1>0$. Since $q=e^{\beta\omega}>1$, this is equivalent to $1<q<2+\sqrt5$, $0<\beta\omega<\ln(2+\sqrt5)$. According to Eq.~\eqref{eq:S_LHS_ss}, we take the parameters $\beta=1$, $\omega=1$, and $\gamma=1$: $\left.\partial_rF^{\rm ss}(\beta;r)\right|_{r=0}\simeq8.72\times10^{-2}>0$. Thus, it satisfies the strict first-order sufficient condition for local steady-state enhancement Eq.~\eqref{eq:S_sscriterion}.

	\section{Transient thermometry and time metrology}
	
	\subsection{Universal short-time distinction}
	We consider diagonal dynamics in a fixed, parameter-independent basis. Let \(\bm p_{\theta,r}(t)\) be the population vector, \(W_\theta\) the classical rate generator induced by the diagonal block of \(\mathcal L_\theta\), and \(\bm p_\delta\) the reset-target population vector. 
	Throughout this subsection we assume full support of the initial population, \((\bm p_0)_i>0\), for all components included in the Fisher-information sum, so that the short-time Fisher-information expansion below is well defined. The reset-modified rate equation is
	\begin{equation}
		\dot{\bm p}_{\theta,r}(t)=W_\theta\bm p_{\theta,r}(t)+r[\bm p_\delta-\bm p_{\theta,r}(t)] ,\label{eq:S_rate_general}
	\end{equation}
	where \(r\), \(\bm p_0\), and \(\bm p_\delta\) are independent of the static parameter \(\theta\). For \(\bm p_{\theta,r}(0)=\bm p_0\),
	\begin{equation}
		\bm p_{\theta,r}(t)=\bm p_0+t\{W_\theta\bm p_0+r(\bm p_\delta-\bm p_0)\}+O(t^2).\label{eq:S_short_population}
	\end{equation}
	For a static parameter \(\theta\),
	\begin{equation}
		\partial_\theta\bm p_{\theta,r}(t)=t(\partial_\theta W_\theta)\bm p_0+O(t^2),\label{eq:S_short_static_derivative}
	\end{equation}
	because the reset term \(r(\bm p_\delta-\bm p_0)\) carries no \(\theta\)-dependence. For diagonal states, the QFI reduces to the classical Fisher information
	\[F_\theta^{(r)}(t)=\sum_i\frac{[\partial_\theta p_{\theta,r,i}(t)]^2}{p_{\theta,r,i}(t)} ,\]
	where \(p_{\theta,r,i}\) denotes the \(i\)-th component of \(\bm p_{\theta,r}\). Using \(p_{\theta,r,i}(t)=(p_0)_i+O(t)\), one obtains
	\begin{equation}
		F_\theta^{(r)}(t)=t^2\sum_i\frac{\bigl[(\partial_\theta W_\theta)\bm p_0\bigr]_i^2}{(\bm p_0)_i}+O(t^3).\label{eq:S_static_short}
	\end{equation}
	Thus reset does not modify the leading \(t^2\) term for static-parameter estimation. Any reset-induced thermometric gain must therefore arise after a finite information-accumulation time.
	The situation is different when the estimated parameter is the physical evolution time itself. The time-estimation Fisher information is
	\begin{equation}
		F_t^{(r)}(t)=\sum_i\frac{\dot p_{\theta,r,i}^2(t)}{p_{\theta,r,i}(t)} .\label{eq:S_time_FI_def}
	\end{equation}
	At \(t=0\),
	\[\dot{\bm p}_{\theta,r}(0)=W_\theta\bm p_0+r(\bm p_\delta-\bm p_0),\]
	and hence
	\begin{equation}
		F_t^{(r)}(0)=\sum_i\frac{\bigl[W_\theta\bm p_0+r(\bm p_\delta-\bm p_0)\bigr]_i^2}{(\bm p_0)_i}.\label{eq:S_time_short}
	\end{equation}
	Equivalently,
	\begin{align}
		F_t^{(r)}(0)=F_t^{(0)}(0)+2r\sum_i\frac{[W_\theta\bm p_0]_i[\bm p_\delta-\bm p_0]_i}{(\bm p_0)_i}+r^2\sum_i\frac{[\bm p_\delta-\bm p_0]_i^2}{(\bm p_0)_i}.\label{eq:S_time_r_expansion}
	\end{align}
	Therefore reset enters the leading short-time drift directly. This explains why time metrology can benefit immediately from reset, whereas static thermometry cannot be enhanced at the leading short-time order.
	
	\subsection{Binary diagonal response coefficient}
	
	For a diagonal state in the same parameter-independent basis,
	\begin{equation}
		\rho_{\theta,r}(t)=\sum_i p_{\theta,r,i}(t)|i\rangle\langle i|,\quad\bm p_{\theta,r}(t)=\{p_{\theta,r,i}(t)\}_i.
	\end{equation}
	The Fisher information is classical:
	\begin{equation}
		F_\theta^{(r)}(t)=\sum_i\frac{[\partial_\theta p_{\theta,r,i}(t)]^2}{p_{\theta,r,i}(t)}.
	\end{equation}
	For a small reset rate,
	\begin{equation}
		p_{\theta,r,i}(t)=p_{\theta,0,i}(t)+r b_{\theta,i}(t)+O(r^2),\quad b_{\theta,i}(t):=\left.\partial_r p_{\theta,r,i}(t)\right|_{r=0}.
	\end{equation}
	Substitution gives
	\begin{equation}
		F_\theta^{(r)}(t)=F_\theta^{(0)}(t)+r\Xi_\theta(t)+O(r^2),
	\end{equation}
	where
	\begin{equation}
		\Xi_\theta(t)=\sum_i\left[\frac{2(\partial_\theta p_{\theta,0,i})(\partial_\theta b_{\theta,i})}{p_{\theta,0,i}}-\frac{(\partial_\theta p_{\theta,0,i})^2 b_{\theta,i}}{p_{\theta,0,i}^2}\right].
		\label{eq:S_Xi_general}
	\end{equation}
	
	For a binary distribution $\{p_{\theta,r}(t),\,1-p_{\theta,r}(t)\}$, define
	\begin{equation}
		p_{\theta,0}(t):=p_{\theta,r}(t)\big|_{r=0},\quad g_\theta(t):=\partial_\theta p_{\theta,0}(t),\quad b_\theta(t):=\left.\partial_r p_{\theta,r}(t)\right|_{r=0}.
	\end{equation}
	Then Eq.~\eqref{eq:S_Xi_general} reduces to
	\begin{equation}
		\Xi_\theta^{\rm bin}(t)=\frac{2g_\theta(t)\partial_\theta b_\theta(t)}{p_{\theta,0}(t)[1-p_{\theta,0}(t)]}-\frac{g_\theta^2(t)b_\theta(t)[1-2p_{\theta,0}(t)]}{p_{\theta,0}^2(t)[1-p_{\theta,0}(t)]^2}.
		\label{eq:S_Xi_binary}
	\end{equation}
	A positive $\Xi_\theta^{\rm bin}(t)$ is the finite-time small-reset sufficient condition for local enhancement.

	\subsection{Thermal-qubit transient thermometry}
	For the thermal qubit used in the Letter, the excited-state population obeys
	\begin{equation}
		\dot p_{T,r}=-(\Gamma+r)p_{T,r}+\Gamma p_{\rm eq}+rp_\delta,
		\label{eq:S_thermal_rate}
	\end{equation}
	where
	\[
	p_{\rm eq}(T)=\frac{1}{1+e^{\omega/T}},\quad
	\Gamma(T)=\gamma[2\bar n(T)+1],\quad
	\bar n(T)=\frac{1}{e^{\omega/T}-1}.
	\]
	During the reset pulse \(0\le t\le t_s\),
	\begin{equation}
		p_{T,r}(t)=p_{T,r}^*+(p_0-p_{T,r}^*)e^{-K_rt},\quad
		K_r=\Gamma+r,\quad
		p_{T,r}^*=\frac{\Gamma p_{\rm eq}+rp_\delta}{\Gamma+r}.
		\label{eq:S_pTr}
	\end{equation}
	Define
	\[
	\chi_r(t):=\partial_Tp_{T,r}(t),\qquad
	\eta_r:=\partial_Tp_{T,r}^* .
	\]
	Differentiating Eq.~\eqref{eq:S_pTr} gives
	\begin{equation}
		\chi_r(t)=\eta_r(1-e^{-K_rt})-(p_0-p_{T,r}^*)t e^{-K_rt}\,\partial_T\Gamma,
		\label{eq:S_chir}
	\end{equation}
	with
	\begin{equation}
		\eta_r=
		\frac{\Gamma(\Gamma+r)\partial_Tp_{\rm eq}+r(p_{\rm eq}-p_\delta)\partial_T\Gamma}{(\Gamma+r)^2},
		\label{eq:S_etar}
	\end{equation}
	and
	\begin{equation}
		\partial_Tp_{\rm eq}=
		\frac{\omega e^{\omega/T}}{T^2(1+e^{\omega/T})^2},\qquad
		\partial_T\Gamma=
		\frac{2\gamma\omega e^{\omega/T}}{T^2(e^{\omega/T}-1)^2}.
		\label{eq:S_dTpeqGamma}
	\end{equation}
	Since the state remains diagonal in the energy basis, the thermometric Fisher information is purely classical:
	\begin{equation}
		F_T^{(r)}(t)=
		\frac{\chi_r^2(t)}
		{p_{T,r}(t)[1-p_{T,r}(t)]},
		\qquad 0\le t\le t_s.
		\label{eq:S_FTr_on}
	\end{equation}
	This is the expression used for the transient-gain map when the reset is kept on until readout, \(t_s=t\).
	
	For \(t>t_s\), the reset is switched off and the system evolves under the original thermalization dynamics:
	\begin{equation}
		p_{T,r}(t)=p_{\rm eq}+[p_{T,r}(t_s)-p_{\rm eq}]e^{-\Gamma(t-t_s)}.
		\label{eq:S_pTr_off}
	\end{equation}
	Its temperature derivative is
	\begin{equation}
		\partial_Tp_{T,r}(t)
		=
		\partial_Tp_{\rm eq}(1-e^{-\Gamma(t-t_s)})
		+
		\bigl[\chi_r(t_s)-(t-t_s)[p_{T,r}(t_s)-p_{\rm eq}]\partial_T\Gamma\bigr]e^{-\Gamma(t-t_s)}.
		\label{eq:S_chir_off}
	\end{equation}
	Hence, after switch-off,
	\begin{equation}
		F_T^{(r)}(t)=
		\frac{[\partial_Tp_{T,r}(t)]^2}
		{p_{T,r}(t)[1-p_{T,r}(t)]},
		\qquad t>t_s.
		\label{eq:S_FTr_off}
	\end{equation}
	Equations~\eqref{eq:S_FTr_on} and \eqref{eq:S_FTr_off} generate the transient thermometry panels.
	
	We next connect these exact finite-\(r\) formulas to the small-reset response. At \(r=0\),
	\begin{equation}
		p(t)=p_{\rm eq}+(p_0-p_{\rm eq})e^{-\Gamma t},
		\label{eq:S_pT0}
	\end{equation}
	and
	\begin{equation}
		g_T(t):=\partial_Tp(t)=
		\partial_Tp_{\rm eq}(1-e^{-\Gamma t})
		-(p_0-p_{\rm eq})t e^{-\Gamma t}\partial_T\Gamma.
		\label{eq:S_gT}
	\end{equation}
	The first-order reset correction is
	\begin{equation}
		b_T(t):=\left.\partial_rp_{T,r}(t)\right|_{r=0}
		=
		\frac{p_\delta-p_{\rm eq}}{\Gamma}(1-e^{-\Gamma t})
		-(p_0-p_{\rm eq})t e^{-\Gamma t}.
		\label{eq:S_bT}
	\end{equation}
	Using the binary response coefficient in Eq.~\eqref{eq:S_Xi_binary}, the local finite-time response is
	\begin{equation}
		\left.\partial_rF_T^{(r)}(t)\right|_{r=0}
		=
		\Xi_T^{\rm bin}(t),
		\label{eq:S_thermo_Xi}
	\end{equation}
	where Eq.~\eqref{eq:S_Xi_binary} is evaluated with
	\[
	p(t)=p_{T,0}(t),\qquad
	g(t)=g_T(t),\qquad
	b(t)=b_T(t),\qquad
	\theta=T.
	\]
	For the parameters :\(\omega=\gamma=1, T=1.5, p_0=0.6, p_\delta=0,\)
	one obtains at \(t=1\)
	\begin{equation}
		\Xi_T^{\rm bin}(1)\simeq3.64\times10^{-2}>0.
		\label{eq:S_XiT_positive}
	\end{equation}
	Thus the plotted finite-time thermometric gain is locally seeded by a positive small-reset response. This does not contradict Eq.~\eqref{eq:S_static_short}: reset is absent from the leading \(t^2\) short-time term, and the thermometric advantage appears only after finite information accumulation.

	\subsection{Thermal-qubit time metrology}
	For the same thermal qubit, the time-estimation Fisher information is obtained from the instantaneous velocity of the excited-state population,
	\[
	F_t^{(r)}(t)=\frac{\dot p_{T,r}^2(t)}
	{p_{T,r}(t)[1-p_{T,r}(t)]}.
	\]
	During the reset pulse, Eq.~\eqref{eq:S_pTr} gives
	\[
	\dot p_{T,r}(t)
	=
	-K_r(p_0-p_{T,r}^*)e^{-K_rt},
	\qquad
	K_r=\Gamma+r.
	\]
	Hence, for \(0\le t\le t_s\),
	\begin{equation}
		F_t^{(r)}(t)
		=
		\frac{K_r^2(p_0-p_{T,r}^*)^2e^{-2K_rt}}
		{p_{T,r}(t)[1-p_{T,r}(t)]}.
		\label{eq:S_Ft_on}
	\end{equation}
	This formula generates the time-metrology gain map when the reset is kept on until readout, \(t_s=t\).
	
	After switch-off, for \(t>t_s\), the population obeys Eq.~\eqref{eq:S_pTr_off}, so one has \(\dot p_{T,r}(t)=-\Gamma[p_{T,r}(t_s)-p_{\rm eq}]e^{-\Gamma(t-t_s)},\)
	and therefore
	\begin{equation}
		F_t^{(r)}(t)
		=
		\frac{\Gamma^2[p_{T,r}(t_s)-p_{\rm eq}]^2e^{-2\Gamma(t-t_s)}}
		{p_{T,r}(t)[1-p_{T,r}(t)]},
		\qquad t>t_s .
		\label{eq:S_Ft_off}
	\end{equation}
	
	The immediate small-reset response follows by evaluating Eq.~\eqref{eq:S_Ft_on} at \(t=0\):
	\begin{equation}
		F_t^{(r)}(0)
		=
		\frac{[\Gamma(p_{\rm eq}-p_0)+r(p_\delta-p_0)]^2}
		{p_0(1-p_0)}.
		\label{eq:S_Ft0}
	\end{equation}
	For \(0<p_0<1\),
	\begin{equation}
		\left.\partial_rF_t^{(r)}(0)\right|_{r=0}
		=
		\frac{2\Gamma(p_{\rm eq}-p_0)(p_\delta-p_0)}
		{p_0(1-p_0)}.
		\label{eq:S_dFt0}
	\end{equation}
	Thus the strict first-order immediate-enhancement condition is
	\begin{equation}
		(p_{\rm eq}-p_0)(p_\delta-p_0)>0.
		\label{eq:S_time_condition}
	\end{equation}
	For the plotted parameters, \(\omega=\gamma=1, T=1.5,p_0=0.6, p_\delta=0,\)
	one has
	\[
	p_{\rm eq}=\frac{1}{1+e^{2/3}}\simeq0.339,
	\qquad
	p_{\rm eq}-p_0\simeq-0.261<0,
	\qquad
	p_\delta-p_0=-0.6<0.
	\]
	Therefore Eq.~\eqref{eq:S_time_condition} is satisfied, and
	\begin{equation}
		\left.\partial_rF_t^{(r)}(0)\right|_{r=0}
		\simeq 4.06>0.
		\label{eq:S_dFt0_positive}
	\end{equation}
	Moreover, Eq.~\eqref{eq:S_Ft0} gives the large-\(r\) short-time scaling
	\begin{equation}
		F_t^{(r)}(0)
		\sim
		r^2
		\frac{(p_\delta-p_0)^2}{p_0(1-p_0)},
		\qquad r\gg\Gamma,
		\label{eq:S_Ft_r2}
	\end{equation}
	which explains why time metrology responds much more directly to reset than thermometry. 
	In contrast to static thermometry, the reset-induced drift enters the leading short-time limit itself.

	\begin{figure}[t]
		\centering
		\includegraphics[width=0.96\linewidth]{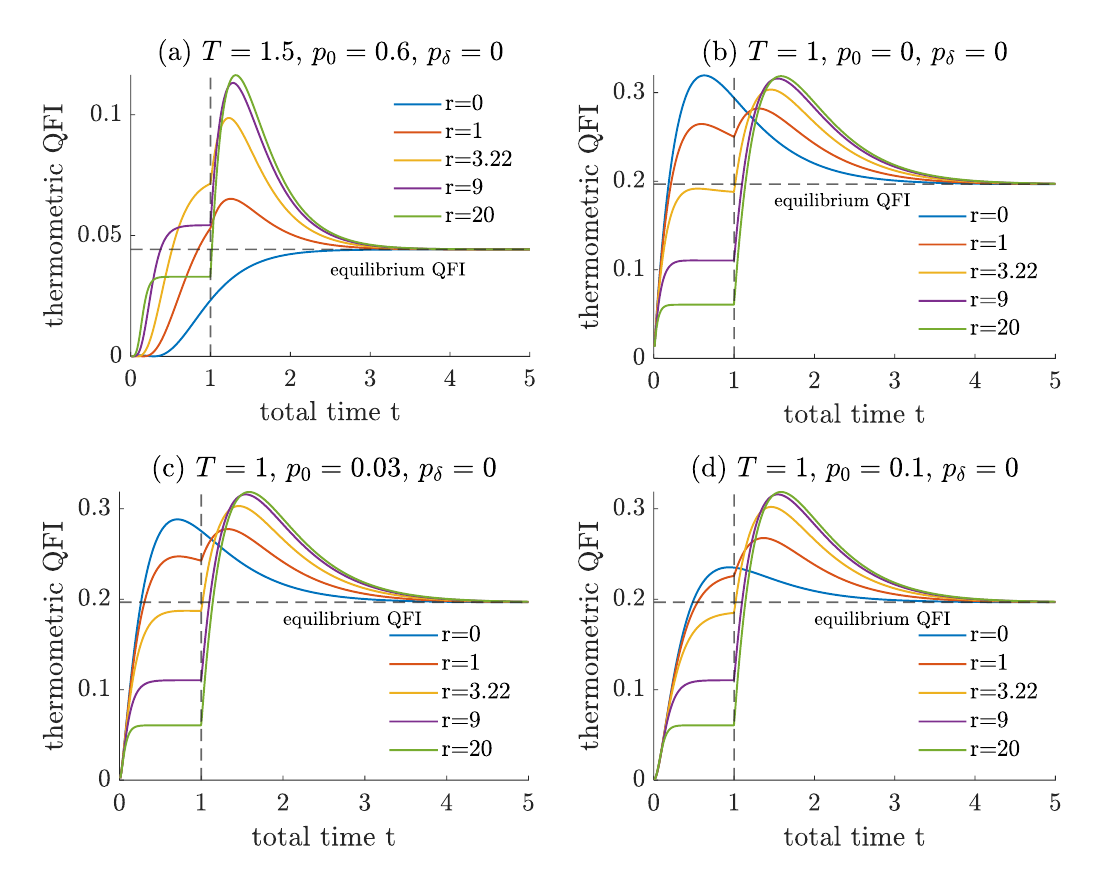}
		\caption{
			Parameter dependence of temporary-reset thermometry. 
			Each panel shows the thermometric Fisher information \(F_T^{(r)}(t)\) for a fixed reset-pulse duration \(t_s=1\), with the dashed vertical line marking switch-off and the dashed horizontal line marking the equilibrium Fisher information. In (a)--(d), \(\omega=\gamma=1\).
			(a) Main-text parameters \(T=1.5\), \(p_0=0.6\), \(p_\delta=0\): temporary reset produces a finite-time thermometric overshoot. 
			(b) \(T=1\), \(p_0=p_\delta=0\): Given that both the initial excited-state population and the reset-target excited-state population are zero, transient thermometric reset cannot be interpreted as a universal peak-enhancement mechanism in this scenario.
			(c,d) Small nonzero initial excited-state populations show the crossover between suppression and finite-time enhancement.	
		}
		\label{fig:S2}
	\end{figure}
	
	\begin{figure}[t]
		\centering
		\includegraphics[width=0.96\linewidth]{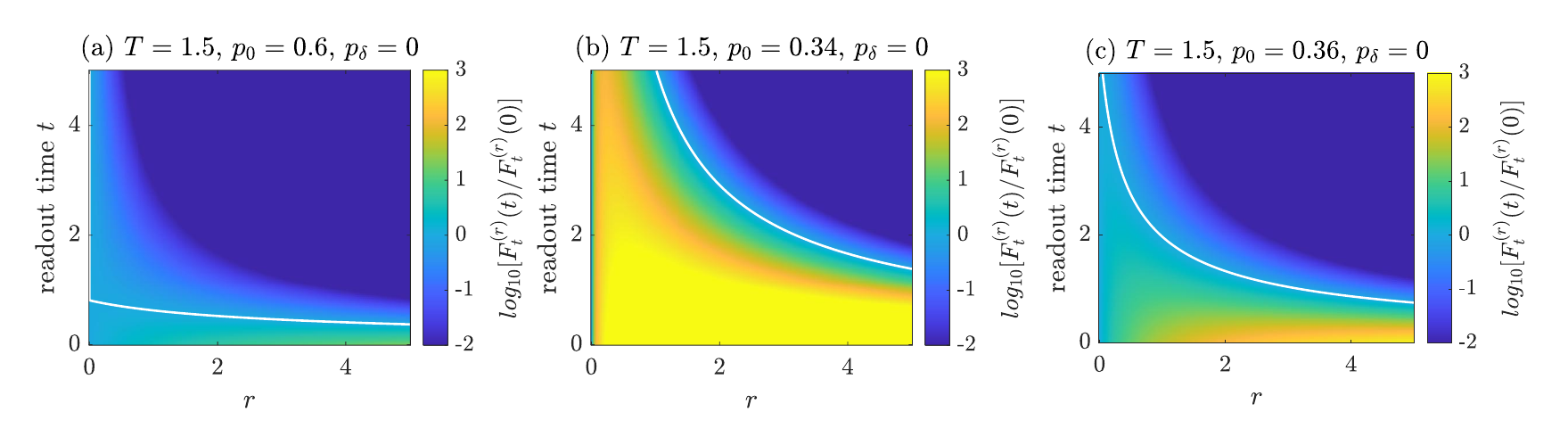}
		\caption{
			Initial-state dependence of transient time-metrology gain. 
			The panels show \(\log_{10}[F_t^{(r)}(t)/F_t^{(0)}(t)]\) as a function of reset rate \(r\) and readout time \(t\). 
			The white curve denotes the unity-gain boundary \(F_t^{(r)}(t)=F_t^{(0)}(t)\). 
			(a) For \(T=1.5\), \(p_0=0.6\), \(p_\delta=0\), reset enhances time metrology at short times, but the gain ratio is moderate because the unreset drift is already finite. 
			(b) For \(p_0=0.34\simeq p_{\rm eq}\), the unreset drift is nearly cancelled, so reset produces a broad and strong gain region. 
			(c) For \(p_0=0.36\), the gain remains visible but is reduced as the initial state moves away from the cancellation point. 
			These results show that the time-metrology enhancement is governed by the reset-induced drift and depends sensitively on the initial excited-state population.
		}
		\label{fig:S3}
	\end{figure}
	
	\subsection{Initial-state dependence of transient reset enhancement}
	
	The transient reset advantage is parameter dependent and is not guaranteed for arbitrary initial excited-state populations. 
	For the thermal qubit, the state remains diagonal in the energy basis, so the QFI reduces to the Fisher information carried by the binary population distribution \(\{p_{T,r}(t),1-p_{T,r}(t)\}\). 
	The finite-time coefficient \(\Xi_T^{\rm bin}(t)\) therefore depends on the initial excited-state population \(p_0\), the reset-target excited-state population \(p_\delta\), the bath temperature \(T\), and the readout time. 
	This dependence is illustrated in Fig.~\ref{fig:S2}. 
	
	For the main parameters \(T=1.5\), \(p_0=0.6\), and \(p_\delta=0\), a temporary reset pulse can move the probe into a high-\(F_T\) window earlier and produce a finite-time overshoot above the equilibrium Fisher information [Fig.~\ref{fig:S2}(a)]. 
	For sufficiently large \(r\), reset pins the state too strongly toward the reset target, which is parameter blind in this thermometry problem, and thereby suppresses the thermal excitation process that carries temperature information during the pulse. 
	
	When \(p_0=p_\delta=0\), the initial excited-state population and the reset-target excited-state population are both zero; equivalently, the qubit starts in the ground state and is reset to the ground state. 
	In this case, reset repeatedly suppresses the thermal excitation dynamics through which temperature information is accumulated. 
	For example, at \(T=1, \omega=\gamma=1, p_0=p_\delta=0,\)
	one finds at \(t=1\),\(\)
	\begin{equation}
		\Xi_T^{\rm bin}(1)\simeq -5.35\times10^{-2}<0 .
	\end{equation}
	Thus small reset locally reduces the thermometric Fisher information, consistent with Fig.~\ref{fig:S2}(b). 
	Figures~\ref{fig:S2}(c,d) show that, for small but nonzero \(p_0\), the response changes continuously as the initial excited-state population is increased. 
	These results confirm that transient thermometric gain is not an unconditional consequence of reset, but a finite-time effect determined by the competition between reset-induced state preparation and thermal information accumulation.
	
	For time metrology, the strict short-time condition Eq.~\eqref{eq:S_time_condition} clarifies the role of \(p_0\). 
	With \(p_\delta=0\), it becomes \((p_{\rm eq}-p_0)(-p_0)>0 .\)
	For \(0<p_0<1\), this is equivalent to \(p_0>p_{\rm eq}.\)
	Thus ground-state reset immediately enhances time metrology only when the initial excited-state population lies above equilibrium. 
	For \(T=1.5\),
	\[
	p_{\rm eq}=\frac{1}{1+e^{2/3}}\simeq0.339 .
	\]
	Therefore \(p_0=0.34\) is already slightly above equilibrium and satisfies the immediate-enhancement condition. 
	This explains the broad bright gain region in Fig.~\ref{fig:S3}(b): near \(p_0\simeq p_{\rm eq}\), the unreset initial drift is close to cancellation, whereas reset introduces the additional drift \(r(p_\delta-p_0)\). 
	Consequently, the gain ratio \(F_t^{(r)}(t)/F_t^{(0)}(t)\) can become very large. 
	This large ratio should be interpreted as a reset-induced lifting of a weak unreset time signal, rather than as a parameter-independent enhancement mechanism.
	
	Figure~\ref{fig:S3} further shows that the time-metrology gain is strongly controlled by the initial excited-state population. 
	For \(p_0=0.6\), the immediate-enhancement condition is satisfied, but the unreset drift is already appreciable, so the gain ratio is moderate. 
	For \(p_0=0.34\simeq p_{\rm eq}\), the unreset baseline is strongly suppressed and reset produces a broad high-gain region. 
	For \(p_0=0.36\), the same mechanism remains visible but becomes weaker as the initial state moves away from the drift-cancellation point. 
	The white curves denote the unity-gain boundary \(F_t^{(r)}(t)=F_t^{(0)}(t)\); the gain region is the side on which
	\[
	\log_{10}\!\left[F_t^{(r)}(t)/F_t^{(0)}(t)\right]>0 .
	\]
	
	If \(p_0=0\), the short-time formula in Eq.~\eqref{eq:S_dFt0} cannot be applied directly because the initial binary distribution is not full rank. 
	This limitation concerns only the strict \(t=0\) time-metrology criterion. 
	For \(p_0=p_\delta=0\), the reset-induced initial drift \(r(p_\delta-p_0)\) vanishes, so the immediate enhancement mechanism captured by Eq.~\eqref{eq:S_time_condition} is absent. 
	Finite-time Fisher information remain well defined for \(t>0\), and their behavior must be evaluated from the full expressions.
	
	\section{Coherent frequency metrology and Bloch decomposition}
	We use the ordered basis \(\{|1\rangle,|0\rangle\}\), so that
	\[
	\sigma_z=
	\begin{pmatrix}
		1&0\\0&-1
	\end{pmatrix},
	\qquad
	\sigma_-=
	|0\rangle\langle1|
	=
	\begin{pmatrix}
		0&0\\1&0
	\end{pmatrix}.
	\]
	Consider a qubit with amplitude damping: $\mathcal L_\omega(\rho) = -i\left[\frac{\omega}{2}\sigma_z, \rho\right] + \gamma\mathcal{D}[\sigma_-]\rho$, $\mathcal{D}[A]\rho = A\rho A^\dagger - \frac{1}{2}\{A^\dagger A, \rho\}.$ 
	Let the excited-state population probability be \( p \), the corresponding density matrix is given by
	\begin{equation}
		\rho_{\omega,r}(t)=\begin{pmatrix}
			p_{\omega,r}(t) & c_{\omega,r}^*(t)\\
			c_{\omega,r}(t) & 1-p_{\omega,r}(t)\end{pmatrix}
		\label{eq:reset_population_matrix}
	\end{equation}
	The Hamiltonian part gives
	\begin{equation}
		-i\left[\frac{\omega}{2}\sigma_z,\rho_{\omega,r}\right]
		=
		\begin{pmatrix}
			0&-i\omega c_{\omega,r}^*\\
			i\omega c_{\omega,r}&0
		\end{pmatrix}.
		\label{eq:S_coh_Hpart}
	\end{equation}
	The amplitude-damping dissipator is
	\begin{align}
		\mathcal D[\sigma_-]\rho_{\omega,r}
		=
		\sigma_-\rho_{\omega,r}\sigma_+
		-
		\frac12\{\sigma_+\sigma_-,\rho_{\omega,r}\}
		=
		\begin{pmatrix}
			-p_{\omega,r}&-\frac12c_{\omega,r}^*\\
			-\frac12c_{\omega,r}&p_{\omega,r}
		\end{pmatrix}.
		\label{eq:S_coh_Dpart}
	\end{align}
	The reset target is
	\[
	\rho_\delta=|+\rangle\langle+|
	=
	\frac12
	\begin{pmatrix}
		1&1\\1&1
	\end{pmatrix},
	\]
	and therefore the reset contribution is
	\begin{equation}
		r(\rho_\delta-\rho_{\omega,r})
		=
		r
		\begin{pmatrix}
			\frac12-p_{\omega,r}&\frac12-c_{\omega,r}^*\\
			\frac12-c_{\omega,r}&p_{\omega,r}-\frac12
		\end{pmatrix}.
		\label{eq:S_coh_Rpart}
	\end{equation}
	Combining Eqs.~\eqref{eq:S_coh_Hpart}--\eqref{eq:S_coh_Rpart}, the reset-period master equation gives the closed scalar equations
	\begin{subequations}
		\begin{align}
			\dot p_{\omega,r}
			&=
			-(\gamma+r)p_{\omega,r}
			+
			\frac{r}{2},
			\label{eq:S_coh_p_ode}\\
			\dot c_{\omega,r}
			&=
			\left(i\omega-\frac{\gamma}{2}-r\right)c_{\omega,r}
			+
			\frac{r}{2}.
			\label{eq:S_coh_c_ode}
		\end{align}
	\end{subequations}
	Here \(p_{\omega,r}(0)=p_0\) and \(c_{\omega,r}(0)=c_0\),
	one finds during the reset pulse
	\begin{subequations}
		\begin{align}
			p_{\omega,r}(t)
			&=
			\frac{r}{2(\gamma+r)}
			+
			\left[
			p_0-\frac{r}{2(\gamma+r)}
			\right]e^{-(\gamma+r)t},
			\label{eq:S_pwr}\\
			c_{\omega,r}(t)
			&=
			\frac{r/2}{\gamma/2+r-i\omega}
			+
			\left[
			c_0-\frac{r/2}{\gamma/2+r-i\omega}
			\right]e^{(i\omega-\gamma/2-r)t}.
			\label{eq:S_cwr}
		\end{align}
	\end{subequations}
	The frequency derivative of the coherence is
	\begin{align}
		\partial_\omega c_{\omega,r}(t)=
		\frac{i r/2}{(\gamma/2+r-i\omega)^2}
		\left[1-e^{(i\omega-\gamma/2-r)t}\right]
		+ it\left[c_0-\frac{r/2}{\gamma/2+r-i\omega}\right]e^{(i\omega-\gamma/2-r)t}.
		\label{eq:S_domegac}
	\end{align}
	Since \(\partial_\omega p_{\omega,r}(t)=0,\)
	any energy-basis population measurement has zero Fisher information for \(\omega\). The frequency information is carried entirely by the coherence.
	
	For a qubit state
	\begin{equation}
		\rho=\frac12(I+\bm s\cdot\bm\sigma),
		\label{eq:S_Bloch_state}
	\end{equation}
	one has explicitly
	\begin{equation}
		\rho=
		\frac12
		\begin{pmatrix}
			1+s_z & s_x-i s_y\\
			s_x+i s_y & 1-s_z
		\end{pmatrix}.
		\label{eq:S_Bloch_matrix}
	\end{equation}
	Thus
	\begin{equation}
		\bm s_{\omega,r}
		=
		\bigl(
		2\operatorname{Re}c_{\omega,r},
		2\operatorname{Im}c_{\omega,r},
		2p_{\omega,r}-1
		\bigr).
		\label{eq:S_blochvector_freq}
	\end{equation}
	
	For completeness, we also recall the qubit QFI formula used below. For Eq.~\eqref{eq:S_Bloch_state}, the SLD can be written as $\mathbb{L}_\theta=a_\theta I+\bm b_\theta\cdot\bm\sigma$. Inserting this into $\partial_\theta\rho=\frac12(\rho \mathbb{L}_\theta+\mathbb{L}_\theta\rho)$ and using $(\bm u\cdot\bm\sigma)(\bm v\cdot\bm\sigma)=(\bm u\cdot\bm v)I+i(\bm u\times\bm v)\cdot\bm\sigma$, one obtains $a_\theta+\bm s\cdot\bm b_\theta=0$, $\bm b_\theta+a_\theta\bm s=\partial_\theta\bm s$.
	Solving these equations gives
	\[
	a_\theta=-\frac{\bm s\cdot\partial_\theta\bm s}{1-|\bm s|^2},
	\qquad
	\bm b_\theta
	=
	\partial_\theta\bm s
	+
	\frac{\bm s\cdot\partial_\theta\bm s}{1-|\bm s|^2}\bm s .
	\]
	Therefore
	\begin{equation}
		F_\theta=\operatorname{Tr}(\rho \mathbb{L}_\theta^2)
		=|\partial_\theta\bm s|^2	+
		\frac{(\bm s\cdot\partial_\theta\bm s)^2}{1-|\bm s|^2}.
		\label{eq:S_qubitQFI}
	\end{equation}
	
	Substituting \( \boldsymbol{s} = s \hat{\boldsymbol{s}} \), we have
	\begin{equation}
		\partial_\theta \boldsymbol{s} = (\partial_\theta s)\hat{\boldsymbol{s}} + s \partial_\theta \hat{\boldsymbol{s}}, \qquad \hat{\boldsymbol{s}} \cdot \partial_\theta \hat{\boldsymbol{s}} = 0,
		\label{eq:deriv_s}
	\end{equation}
	Thus
	\begin{equation}
		\|\partial_\theta \boldsymbol{s}\|^2 = (\partial_\theta s)^2 + s^2 \|\partial_\theta \hat{\boldsymbol{s}}\|^2, \qquad \boldsymbol{s} \cdot \partial_\theta \boldsymbol{s} = s \partial_\theta s.
		\label{eq:norm_decomp}
	\end{equation}
	Substituting back into Eq.~\eqref{eq:deriv_s} and \eqref{eq:norm_decomp} into Eq.~\eqref{eq:S_qubitQFI}, we obtain
	\begin{equation}
		F_\theta = \frac{(\partial_\theta s)^2}{1 - s^2} + s^2 \|\partial_\theta \hat{\boldsymbol{s}}\|^2.
		\label{eq:Ftheta_final}
	\end{equation}
	This is precisely the classical-quantum decomposition of the Bloch vector: the first term corresponds to changes in the density-matrix eigenvalue, and the second term to changes in the eigenvector contributions.
	
	For frequency estimation, \(\partial_\omega p_{\omega,r}=0\). 
	Using Eq.~\eqref{eq:S_blochvector_freq}, we have
	\begin{equation}
		|\partial_\omega\bm s_{\omega,r}|^2
		=
		4|\partial_\omega c_{\omega,r}|^2,
		\label{eq:S_bloch_norm}
	\end{equation}
	\begin{equation}
		\bm s_{\omega,r}\cdot\partial_\omega\bm s_{\omega,r}
		=
		4\operatorname{Re}
		\left[
		c_{\omega,r}^*\partial_\omega c_{\omega,r}
		\right]
		=
		2\partial_\omega |c_{\omega,r}|^2 .
		\label{eq:S_bloch_dot}
	\end{equation}
	\begin{equation}
		1-|\bm s_{\omega,r}|^2
		=
		4\left[
		p_{\omega,r}(1-p_{\omega,r})-|c_{\omega,r}|^2
		\right].
		\label{eq:S_bloch_den}
	\end{equation}
	Substituting Eqs.~\eqref{eq:S_bloch_norm}--\eqref{eq:S_bloch_den} into Eq.~\eqref{eq:S_qubitQFI} gives the exact QFI
	\begin{equation}
		F_\omega^{(r)}(t)
		=
		4|\partial_\omega c_{\omega,r}|^2
		+
		\frac{
			[\partial_\omega |c_{\omega,r}|^2]^2
		}{
			p_{\omega,r}(1-p_{\omega,r})-|c_{\omega,r}|^2
		}.
		\label{eq:S_Fomega}
	\end{equation}
	The expression above applies directly to mixed states with
	\(p_{\omega,r}(1-p_{\omega,r})-|c_{\omega,r}|^2>0\).
	For a pure-state trajectory, the corresponding branch is
	\(F_\omega=4|\partial_\omega c_{\omega,r}|^2\), since
	\(\partial_\omega p_{\omega,r}=0\) and purity implies
	\(\partial_\omega|c_{\omega,r}|^2=0\).
	At \(r=0\), \(	c_{\omega,0}(t)=c_0e^{(i\omega-\gamma/2)t},
	\partial_\omega c_{\omega,0}(t)
	=itc_0e^{(i\omega-\gamma/2)t},\)
	and \(\partial_\omega|c_{\omega,0}(t)|^2=0\). Hence
	\begin{equation}
		F_\omega^{(0)}(t)
		=
		4|c_0|^2t^2e^{-\gamma t}.
		\label{eq:S_Fomega0}
	\end{equation}
	
	\subsection{Verification of the QFI-response identity}
	
	Let \(\rho_{\omega,r}(t)=\frac12[I+\bm s_{\omega,r}(t)\cdot\bm\sigma],
	\bm u=\partial_\omega\bm s,\bm v=\partial_r\bm s,\)
	and define \(	M=\bm s\cdot\bm u,	G=1-|\bm s|^2.\)
	The SLD is
	\begin{equation}
		\mathbb{L}_{\omega,r}
		=-\frac{M}{G}I	+
		\left(
		\bm u+\frac{M}{G}\bm s
		\right)\cdot\bm\sigma.
		\label{eq:S_BlochSLD}
	\end{equation}
	The QFI is
	\begin{equation}
		F_\omega=|\bm u|^2+\frac{M^2}{G}.
		\label{eq:S_BlochF}
	\end{equation}
	Differentiating Eq.~\eqref{eq:S_BlochF} with respect to \(r\) gives
	\begin{equation}
		\partial_rF_\omega
		=
		2\bm u\cdot\partial_r\bm u
		+
		\frac{2M}{G}
		\left(
		\bm v\cdot\bm u+\bm s\cdot\partial_r\bm u
		\right)
		+
		\frac{2M^2}{G^2}\bm s\cdot\bm v.
		\label{eq:S_Bloch_dF}
	\end{equation}
	
	A direct trace evaluation gives the same expression. Let $\bm u_r:=\partial_r\bm u=\partial_r\partial_\omega\bm s$, $\mathbb{L}_{\omega,r}=aI+\bm b\cdot\bm\sigma$, where, from Eq.~\eqref{eq:S_BlochSLD}, $a=-\frac{M}{G}$, $\bm b=\bm u+\frac{M}{G}\bm s$. 
	Using $\partial_r\rho_{\omega,r}=\frac12\bm v\cdot\bm\sigma$, $\partial_\omega\partial_r\rho_{\omega,r}=\frac12\bm u_r\cdot\bm\sigma$, and the Pauli trace identities $\Tr(\sigma_i)=0$, $\Tr(\sigma_i\sigma_j)=2\delta_{ij}$,
	the first term becomes
	\begin{equation}
		2\Tr[(\partial_\omega\partial_r\rho_{\omega,r})\mathbb{L}_{\omega,r}]=2\Tr\left[\frac12(\bm u_r\cdot\bm\sigma)(aI+\bm b\cdot\bm\sigma)\right]=2\bm u_r\cdot\bm b.
		\label{eq:S_response_term1}
	\end{equation}
	For the second term, use
	\[
	\mathbb{L}_{\omega,r}^2=(aI+\bm b\cdot\bm\sigma)^2=(a^2+|\bm b|^2)I+2a\,\bm b\cdot\bm\sigma,
	\]
	where $(\bm b\cdot\bm\sigma)^2=|\bm b|^2I$. Hence
	\begin{equation}
		\Tr[(\partial_r\rho_{\omega,r})\mathbb{L}_{\omega,r}^2]=\Tr\left[\frac12(\bm v\cdot\bm\sigma)\left((a^2+|\bm b|^2)I+2a\,\bm b\cdot\bm\sigma\right)\right]=2a\,\bm v\cdot\bm b.
		\label{eq:S_response_term2}
	\end{equation}
	Therefore the right-hand side of the response identity is
	\begin{align}
		&2\Tr[(\partial_\omega\partial_r\rho_{\omega,r})\mathbb{L}_{\omega,r}]-\Tr[(\partial_r\rho_{\omega,r})\mathbb{L}_{\omega,r}^2]
		=2\bm u_r\cdot\bm b-2a\,\bm v\cdot\bm b
		=2\bm u\cdot\partial_r\bm u+\frac{2M}{G}\left(\bm v\cdot\bm u+\bm s\cdot\partial_r\bm u\right)+\frac{2M^2}{G^2}\bm s\cdot\bm v.
		\label{eq:S_Bloch_response_equal}
	\end{align}
	Thus \(\partial_rF_\omega[\rho_{\omega,r}(t)]
	=2\Tr[(\partial_\omega\partial_r\rho_{\omega,r})\mathbb{L}_{\omega,r}]-
	\Tr[(\partial_r\rho_{\omega,r})\mathbb{L}_{\omega,r}^2],\)
	so the coherent frequency trajectory satisfies the general QFI-response identity identically.
	
	\subsection{First-order frequency response and positivity}
	
	We now verify that the coherent frequency-metrology example used in the Letter is locally enhanced by a small reset rate. 
	From Eq.~\eqref{eq:S_Fomega}, define
	\[
	A_\omega:=\partial_\omega c_{\omega,r},\qquad
	B_\omega:=\partial_\omega |c_{\omega,r}|^2
	=2\operatorname{Re}(c_{\omega,r}^*A_\omega),
	\qquad
	D_\omega:=p_{\omega,r}(1-p_{\omega,r})-|c_{\omega,r}|^2 .
	\]
	Then
	\[
	F_\omega^{(r)}(t)=4|A_\omega|^2+\frac{B_\omega^2}{D_\omega}.
	\]
	Differentiating with respect to \(r\) gives
	\begin{equation}
		\partial_rF_\omega^{(r)}
		=
		8\operatorname{Re}(A_\omega^*\partial_rA_\omega)
		+
		\frac{2B_\omega\partial_rB_\omega}{D_\omega}
		-
		\frac{B_\omega^2\partial_rD_\omega}{D_\omega^2}.
		\label{eq:S_dFomega}
	\end{equation}
	
	At \(r=0\),
	\[
	c_{\omega,0}(t)=c_0e^{(i\omega-\gamma/2)t},
	\qquad
	A_{\omega,0}(t)=itc_0e^{(i\omega-\gamma/2)t},
	\qquad
	p_{\omega,0}(t)=p_0e^{-\gamma t}.
	\]
	The first-order reset corrections are
	\begin{subequations}
		\begin{align}
			p_1(t)
			&:=
			\left.\partial_rp_{\omega,r}(t)\right|_{r=0}
			=
			\frac{1-e^{-\gamma t}}{2\gamma}
			-
			p_0t e^{-\gamma t},
			\label{eq:S_p1}\\
			c_1(t)
			&:=
			\left.\partial_rc_{\omega,r}(t)\right|_{r=0}
			=
			\frac{1-e^{(i\omega-\gamma/2)t}}{\gamma-2i\omega}
			-
			c_0t e^{(i\omega-\gamma/2)t},
			\label{eq:S_c1}\\
			A_1(t)
			&:=\partial_\omega c_1(t)=\partial_\omega \left.\partial_rc_{\omega,r}(t)\right|_{r=0}
			=
			\frac{
				-it e^{(i\omega-\gamma/2)t}(\gamma-2i\omega)
				+
				2i[1-e^{(i\omega-\gamma/2)t}]
			}
			{(\gamma-2i\omega)^2}
			-
			ic_0t^2e^{(i\omega-\gamma/2)t}.
			\label{eq:S_A1}
		\end{align}
	\end{subequations}
	Therefore
	\begin{equation}
		\left.\partial_rF_\omega^{(r)}(t)\right|_{r=0}
		=
		\Xi_\omega(t),
	\end{equation}
	where
	\begin{equation}
		\Xi_\omega(t)
		=
		8\operatorname{Re}(A_{\omega,0}^*A_1)
		+
		\frac{2B_0B_1}{D_0}
		-
		\frac{B_0^2D_1}{D_0^2}.
		\label{eq:S_Xiomega}
	\end{equation}
	Here
	\[
	B_0=2\operatorname{Re}(c_{\omega,0}^*A_{\omega,0}),
	\qquad
	B_1=2\operatorname{Re}(c_1^*A_{\omega,0}+c_{\omega,0}^*A_1),
	\]
	and
	\[
	D_0=p_{\omega,0}(1-p_{\omega,0})-|c_{\omega,0}|^2,
	\qquad
	D_1=(1-2p_{\omega,0})p_1-2\operatorname{Re}(c_{\omega,0}^*c_1).
	\]
	For the plotted parameters
	\[
	\gamma=\omega=1,\qquad
	p_0=0.9,\qquad
	c_0=0.075,\qquad
	t=1,
	\]
	we obtain
	\begin{equation}
		F_\omega^{(0)}(1)\simeq8.28\times10^{-3},
		\qquad
		\Xi_\omega(1)\simeq4.31\times10^{-2}>0.
		\label{eq:S_Xiomega_positive}
	\end{equation}
	Thus the coherent frequency-metrology gain shown in the Letter is locally seeded by a positive first-order reset response.

	\subsection{Bloch decomposition into eigenvalue and eigenvector contributions}
	
	Using Eq.~\eqref{eq:S_qubitQFI}, write
	\[
	\bm s=s\hat{\bm s},
	\qquad
	s=|\bm s|,
	\qquad
	|\hat{\bm s}|=1 .
	\]
	Since
	\[
	\partial_\theta\bm s
	=
	(\partial_\theta s)\hat{\bm s}
	+
	s\partial_\theta\hat{\bm s},
	\qquad
	\hat{\bm s}\cdot\partial_\theta\hat{\bm s}=0,
	\]
	we have
	\[
	|\partial_\theta\bm s|^2
	=
	(\partial_\theta s)^2
	+
	s^2|\partial_\theta\hat{\bm s}|^2,
	\qquad
	\bm s\cdot\partial_\theta\bm s
	=
	s\partial_\theta s .
	\]
	Substituting these identities into Eq.~\eqref{eq:S_qubitQFI} yields
	\begin{equation}
		F_\theta
		=
		\frac{(\partial_\theta s)^2}{1-s^2}
		+
		s^2|\partial_\theta\hat{\bm s}|^2
		\equiv
		F_\theta^{\rm classical}
		+
		F_\theta^{\rm quantum}.
		\label{eq:S_BlochSplit}
	\end{equation}
	The first term is the eigenvalue contribution, because the eigenvalues of \(\rho\) are \((1\pm s)/2\). 
	The second term is the eigenvector-rotation contribution, because it depends on the change of the Bloch-vector direction \(\hat{\bm s}\). 
	In the coherent frequency example, the optimized gain is dominated by \(F_\omega^{\rm quantum}\), confirming that the reset advantage is coherence driven rather than population driven.

	\section{Remarks on local and finite-reset advantage}
	
	The conditions derived above are local small-reset conditions. Positivity of $\partial_rF|_{r=0}$ guarantees that $F(r)>F(0)$ for sufficiently small positive $r$, but it is not equivalent to global finite-$r$ optimality. The steady-state thermometry example is special because the exact finite-$r$ inequality can be solved analytically and yields the same temperature boundary for the existence of an advantage. In transient thermometry and frequency metrology, finite-$r$ gain maps may contain broader optimal regions, whereas the local coefficients $\Xi_T(t)$, $\partial_rF_t^{(r)}(0)|_{r=0}$, and $\Xi_\omega(t)$ verify that the plotted improvements are consistent with the first-order response theory.

	% reference
	% No bibliography database is needed in this supplement.